\documentclass[twocolumn,pra, aps,superscriptaddress]{revtex4-1}

\usepackage{mathptmx}
\usepackage{subfigure}
\usepackage{dcolumn}
\usepackage{amsmath,amssymb}
\usepackage{bm}
\usepackage{color}
\usepackage{overpic}
\usepackage{latexsym}
\usepackage{epstopdf}
\usepackage{color}
\usepackage[english]{babel}

\usepackage{psfrag,graphicx} 
\usepackage{epsf} 
\usepackage{subfigure} 
\usepackage{amsfonts}
\usepackage{bm}
\usepackage{natbib}
\usepackage{epstopdf}
\usepackage{appendix}

\definecolor{mygrey}{gray}{0.35}
\definecolor{myblue}{rgb}{0.2,0.2,0.8}
\definecolor{myzard}{cmyk}{0,0,0.05,0}
\definecolor{mywhite}{rgb}{1,1,1}
\definecolor{myred}{rgb}{1,0.,0.3}

\usepackage[colorlinks=true,citecolor=myblue,linkcolor=myred]{hyperref}
\def\be{\begin{equation}}
\def\ee{\end{equation}}
\def\ba{\begin{align}}
\def\enda{\end{align}}
\def\bi{\begin{itemize}}
\def\ei{\end{itemize}}

 \def\ee{\mathord{\rm e}}
 
 \def\ii{\mathord{\rm i}}

\def\half{\textstyle\frac{1}{2}}

 \def\ee{\mathord{\rm e}}
 
 \def\ii{\mathord{\rm i}}

\def\half{\textstyle\frac{1}{2}}

\renewcommand{\ii}{{\rm i}}
\renewcommand{\ee}{{\rm e}}

\def\beq{\begin{equation}}
\def\beq{\begin{equation}}
\def\eeq{\end{equation}}

 \newcommand{\ket}[1]{|#1\rangle}
 \newcommand{\bra}[1]{\langle #1|}

 \newcommand{\ketbradif}[2]{\ket{#1}\bra{#2}}
 \newcommand{\ketbra}[1]{\ketbradif {#1}{#1}}

\begin{document}


\title[Short Title]{Dissipative ground-state preparation of a spin chain by a structured environment}

\author{Cecilia Cormick}
\affiliation{Institute for Theoretical Physics, Universit\"at Ulm, D 89069 Ulm, Germany\\Center for Integrated Quantum
Science and Technology, Universit\"at Ulm, 89069 Ulm, Germany}
\author{Alejandro Bermudez}
\affiliation{Institute for Theoretical Physics, Universit\"at Ulm, D 89069 Ulm, Germany\\Center for Integrated Quantum
Science and Technology, Universit\"at Ulm, 89069 Ulm, Germany}
\author{Susana F. Huelga}
\affiliation{Institute for Theoretical Physics, Universit\"at Ulm, D 89069 Ulm, Germany\\Center for Integrated Quantum
Science and Technology, Universit\"at Ulm, 89069 Ulm, Germany}
\author{Martin B. Plenio}
\affiliation{Institute for Theoretical Physics, Universit\"at Ulm, D 89069 Ulm, Germany\\Center for Integrated Quantum
Science and Technology, Universit\"at Ulm, 89069 Ulm, Germany}



\begin{abstract}

We propose a dissipative method to prepare the ground state of the isotropic
XY spin Hamiltonian in a transverse field. Our model consists of a spin chain with nearest-neighbour
interactions and an additional collective coupling of the spins to a damped harmonic oscillator. The latter provides an
effective environment with a Lorentzian spectral density and can be used to drive the chain asymptotically towards its
multipartite-entangled ground state at a rate that depends on the degree of non-Markovianity of the evolution. We also
present a detailed proposal for the experimental implementation with a
chain of trapped ions. The protocol does not require individual addressing, concatenated pulses, or
multi-particle jump operators, and is capable of generating the desired target state in small
ion chains with very high fidelities. 

\end{abstract}

\maketitle

\section{Introduction}

The general goal of quantum information processing is to manipulate the information coded in a particular quantum
system, while simultaneously trying to isolate it from undesired perturbations due to the external environment. For
instance, in the fields of quantum
computation~\cite{qc_review} and quantum simulation~\cite{qs_review}, the processing stage usually relies on a designed
unitary evolution which, preserving the coherence of the system, provides a  quantum state that contains the outcome of
the computation, or the properties of the quantum phase of matter under study. Therefore, the experimental progress in
these fields has been typically associated to  technological developments that minimize experimental imperfections and
maximize environmental isolation~\cite{qc_review,qs_review}.

However, the coupling of a system to its surrounding environment is not necessarily a disadvantage. In fact, the
dissipation of a quantum system, if judiciously exploited, can act as a resource for quantum information processing.
This interesting change of paradigm started with the recognition that dissipation can assist the generation of
entanglement between distant atoms, either in free space~\cite{entanglement_two_atoms_free_space}, or trapped inside
cavities~\cite{entanglement_same_cavity_no_photon_detection,entanglement_two_atoms_cavities}. By measuring the presence
of a photon  spontaneously emitted by the
atoms~\cite{entanglement_two_atoms_free_space,entanglement_two_atoms_cavities}, or its
absence~\cite{entanglement_same_cavity_no_photon_detection}, it is possible to project an initially uncorrelated atomic
state into a maximally entangled one, as has been recently experimentally
demonstrated~\cite{entanglement_two_atoms_experiment}. These schemes can be improved further by tailoring the
atom-cavity
interaction~\cite{entanglement_two_atoms_cavities_improvement} or by modifying the 
measurement techniques~\cite{entanglement_same_cavity_homodyne}, and can also be generalized to provide a route towards
universal quantum computation~\cite{no_photon_dfs_computation}. Since the dissipation does not always render the
desired result in these approaches, the measurement is required  to select only the successful outcomes. 
Therefore, the combination of dissipation and measurement can be considered as a {probabilistic} resource for quantum
information.

Another promising approach is to assess whether, by designing the system-environment coupling in a particular way,  the
dissipation could provide the desired quantum state with certainty. This {\it quantum reservoir engineering} finds its
roots in the theory of laser cooling of atoms~\cite{metcalf} and, more related to this work, of trapped
ions~\cite{laser_cooling_ti_review}. Laser-cooling schemes try to design the
system-environment coupling such that the dissipation yields a stationary state with reduced kinetic energy. These ideas
can be taken a step further in order to non-probabilistically produce stationary states that display non-classical
aspects, or a certain amount of 
entanglement~\cite{reservoir_engineering}. For instance, by mimicking superradiance
phenomena~\cite{dicke}, certain engineered decay channels yield partial entanglement in the stationary mixed
state~\cite{superradiance_channels}. 
However, it would be highly desirable to devise dissipative protocols that provide {\it maximally-entangled pure states}
asymptotically, the so-called dark states, showing fidelities comparable to those obtained by more standard unitary
schemes~\cite{qc_review}. 

This approach has been recently pursued by different groups which have shown that, provided that one can engineer a
dissipation that acts {quasi-locally} on different parts of the system, it is possible to design dissipative protocols
to produce a number of paradigmatic dark states~\cite{dissipative_pure_state_engineering}, or  to perform universal
dissipative quantum computation~\cite{dissipative_qc}. For instance, by engineering a particular dissipation that acts
on pairs of adjacent atoms in an optical lattice, the system can be driven dissipatively into a superfluid
phase~\cite{driven_dissipative_condensate}. In addition, as initially proposed for Rydberg
atoms~\cite{rydberg_dissipation} and first realized in experiments of trapped ions~\cite{digital_dissipation_ions}, by
concatenating  multi-qubit gates with a controlled dissipation of ancillary qubits, a variety of multipartite entangled
states (i.e. Bell, Greenberger-Horne-Zeilinger, and Dicke states) have been dissipatively generated  with 
reasonably high fidelities. Let us note that this stroboscopic time-evolution corresponds to a Markovian master equation
in the limit of many gates and dissipation steps, where the dissipative jump operators correspond to multi-qubit jump
operators. From a fundamental point of view, it is of interest to address whether similar dissipative protocols
can also work in an {\it (i) analog} and {\it (ii) global} fashion (i.e. using always-on couplings without individual
addressing). Additionally, from a more pragmatic point of view, such global analog schemes would not be limited by the
accumulation of the errors in each step of the stroboscopic protocols. However, finding particular schemes to  provide 
multi-qubit jump operators in an analog manner seems to be a tremendous task from both a theoretical and experimental
point of view. Therefore, we impose a further constraint, the jump operators should be composed of {\it (iii)
single-qubit operators}.

In this article, we develop an instance of a global analog dissipative protocol that generates multipartite entangled
states corresponding to ground states of a quantum spin chain. The underlying idea is to engineer jump operators that
are a particular sum of single-spin operators, in analogy to the models of collective spontaneous emission~\cite{dicke}.
We show that, if the dissipation of the spins is mediated by a common harmonic mode, the spin chain sees a
structured Markovian environment. By structured environment we refer to a spectral density exhibiting a
sharp peak at a certain frequency (so called Breit-Wigner resonance), corresponding to a weakly damped harmonic
oscillator. This structure will allow us to design jump operators in such a way that the
stationary state of the system corresponds to the ground state of the quantum spin chain. In particular,
we consider an isotropic XY spin chain, which describes a critical phase of matter. 
 
We explain in detail how to implement this protocol with trapped ions, relying on tools
that have already been achieved experimentally.  These tools are the so-called state-dependent
forces~\cite{state_dependent_forces_cats,state_dependent_forces_gates,state_dependent_forces_simulations,
state_dependent_forces_walks,state_dependent_forces_mw}, and sympathetic resolved-sideband cooling developed for quantum
computation~\cite{symp_sideband_cooling, sympathetic_cooling}. We test the scheme for realistic parameters, and show 
that the fidelities that
can be achieved are comparable to the unitary protocols that produce multipartite entangled states
~\cite{entangled_states_ions}. Our procedure performs well for small chains of trapped ions, which is
actually the regime in most of the current experiments. 
 
This article is organized as follows. In Sec.~\ref{dissipative_protocol}, we describe the dissipative protocol. We start
by introducing the model under consideration in Sec.~\ref{model}, and then move onto an analytic discussion, supported
by
numerical results, of how to engineer a structured environment that allows us to prepare multipartite entangled states
dissipatively (Sec.~\ref{cooling_spin_wave}). In Sec.~\ref{spin_gs_cooling} we explain how the method can be used to
cool
the system to the many-body ground state of the quantum spin chain. In Sec.~\ref{sec:ions} we show how crystals of
trapped atomic ions are ideally suited for an experimental implementation of our ideas. Section~\ref{ion_crystal} 
describes the trapped-ion setup, a two-species Coulomb crystal. In
Secs.~\ref{ion_xy_model},~\ref{ion_spin_boson} and~\ref{ion_damping}  we introduce the trapped-ion toolbox
required to implement the dissipative protocol. Section~\ref{numerical} contains a numerical analysis of the 
trapped-ion dissipative protocol. Finally, we present our conclusions and an outlook in Sec.~\ref{conclusions}.
 
\section{Steady-state entanglement of a spin chain}
\label{dissipative_protocol}

\subsection{Spin chain with controllable decoherence}
\label{model}

\begin{figure}

\centering
\includegraphics[width=1\columnwidth]{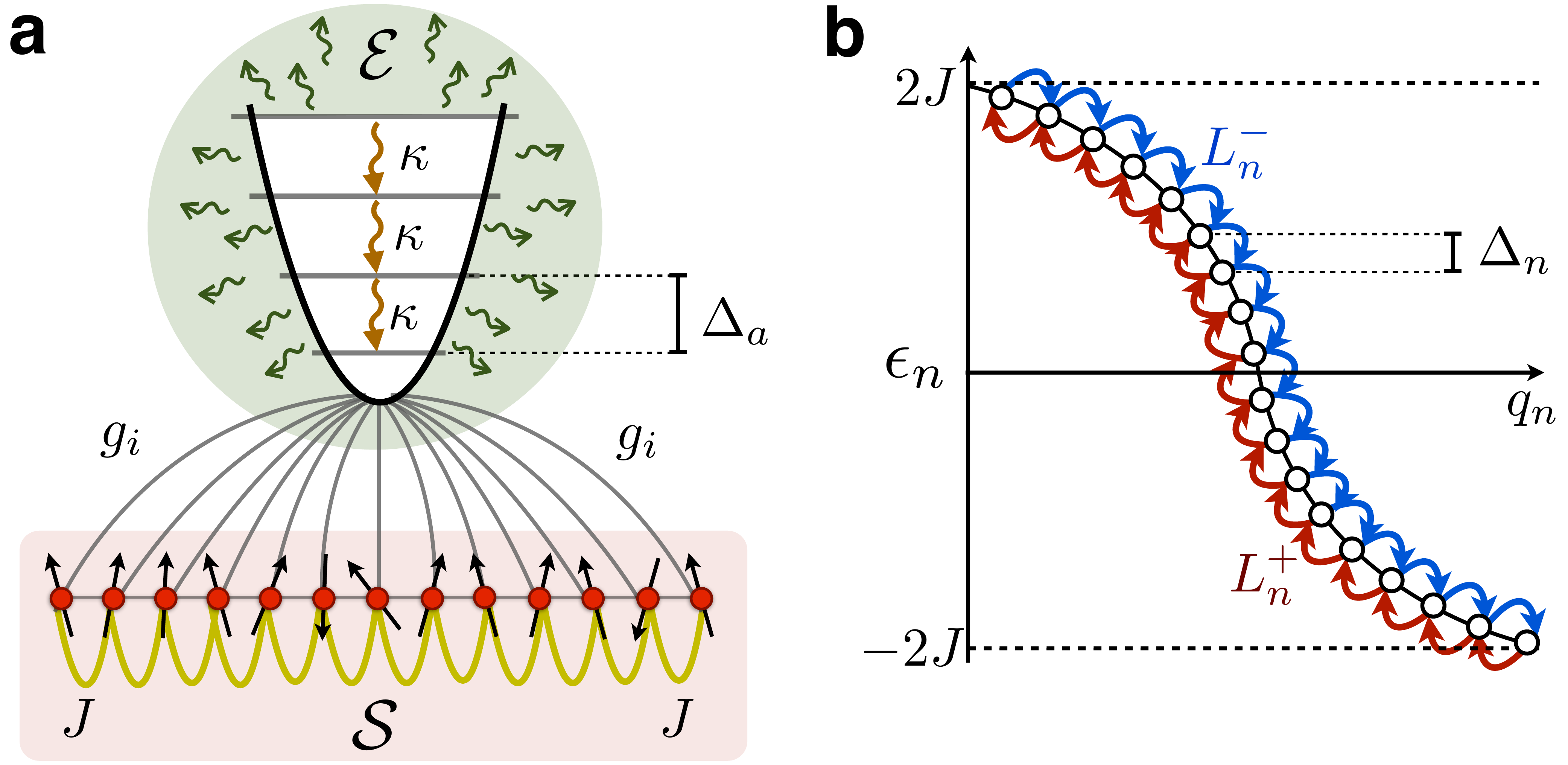}
\caption{ {\bf Damped spin-boson chain:} {\bf (a)} Schematic representation of a spin chain with an isotropic XY
Hamiltonian subjected to an additional transverse field whose intensity depends on the position coordinate of
a damped harmonic oscillator. This oscillator mediates the dissipation of the spin chain, which effectively sees a
structured environment. {\bf (b)} The effect of the oscillator on the spin chain is described by jump operators
$L_n^+$ and $L_n^-$ going respectively up and down the spectrum of spin-chain excitations with quasi-momentum $q_n$ and
energy $\epsilon_n$ (the model parameters are defined in the text).}
\label{fig_scheme}
\end{figure}

Let us start by introducing the model under consideration: an interacting spin-$\half$ chain coupled to a damped bosonic
mode (see Fig.~\ref{fig_scheme}{\bf (a)}). This model, which  shall be referred to as the {\it damped spin-boson chain}
(DSBC) in the rest of the manuscript, is a many-body generalization of a system considered recently for the
dissipative generation of two-qubit entanglement~\cite{non_markovianity_2_spins}. The DSBC is
 described by the following  master equation ($\hbar=1$)
\begin{equation}
\label{dsbc}
\frac{{\rm d}\rho}{{\rm d}t}=\mathcal{L}_{\rm DSBC}(\rho)=-\ii[H_{\rm b}+H_{\rm s}+H_{\rm sb},\rho]+\mathcal{D}_{\rm
b}(\rho),
\end{equation}
where $\rho$ is the  total density matrix, and $\mathcal{L}_{\rm DSBC}$ is the dissipative Liouvillian. We consider a
finite chain of $N$ spins evolving under  the  { isotropic XY Hamiltonian}~\cite{xy_model} 
\begin{equation}
\label{xy_model}
H_{\rm s}=\sum_{i=1}^{N-1}J\sigma_i^+\sigma_{i+1}^-+\text{H.c.},
\end{equation}
where $J\!>\!0$ is an antiferromagnetic interaction strength, and we introduced the raising and lowering operators
$\sigma_i^+=\ket{{\uparrow}_i}\bra{{\downarrow}_i}=(\sigma_i^-)^{\dagger}$. The bosonic Hamiltonian consists of a single
mode
\begin{equation}
\label{oscillator}
H_{\rm b}=\Delta_aa^{\dagger}a,
\end{equation}  
 where $a^{\dagger}$ and $a$ are respectively the creation and annihilation operators (the oscillator
frequency $\Delta_a$ can actually correspond to a detuning $\Delta_a\lessgtr0$, as will become clear from the
experimental
implementation).  The spins and the boson are coupled through
\begin{equation}
\label{spin_boson}
H_{\rm sb}=\sum_{i=1}^{N}g_i\sigma_i^z(a+a^{\dagger}).
\end{equation}  
In the above expression, we have introduced the  site-dependent  coupling strength $g_i$, and the Pauli matrices
$\sigma_i^z=\ket{{\uparrow}_i}\bra{{\uparrow}_i}-\ket{{\downarrow}_i}\bra{{\downarrow}_i}$. Finally, we also consider 
weak damping of the bosonic mode via a Lindblad-type dissipator 
\begin{equation}
\label{damping}
\mathcal{D}_{\rm b}(\rho)=\kappa\left(a\rho a^{\dagger}-a^{\dagger}a\rho\right)+\text{H.c.},
\end{equation}
where $\kappa>0$ is the damping rate. We note that all these individual ingredients can be realized with the
state-of-the-art technology of trapped-ion crystals, as described in Sec.~\ref{sec:ions}. The objective of this work is
to understand the interplay of these dynamics, such that the  Liouvillian $\mathcal{L}_{\rm DSBC}$ generates
stationary entanglement between the initially uncorrelated spins. So far, we can already appreciate two of the
properties of the protocol outlined in the introduction: {\it (i)} It is {\it analog}, since the
couplings~\eqref{xy_model}-\eqref{damping}  are considered to be switched on during the whole protocol. {\it (ii)} It is
{\it global}, since the couplings~\eqref{xy_model}-\eqref{damping} address all the spins of the chain.

Let us note that the  spin-chain Hamiltonian~\eqref{xy_model} alone could already generate entangled states unitarily at
given instants of time. Nonetheless, these correlations have a transient nature, whereas the interest of this work is 
the onset of stationary  correlations. Therefore, we  explore the interplay of this  Hamiltonian part with the
irreversible dynamics introduced by the damped boson through the spin-boson  coupling~\eqref{spin_boson}. 

In Fig.~\ref{fig_scheme}{\bf (a)}, we represent schematically the DSBC. The spin chain is considered to be our system
$\mathcal{S}$, whereas the bosonic mode, together with the Markovian reservoir where it dissipates, provide an
effective structured environment $\mathcal{E}$ for the spin
chain. At first sight, the dephasing-like spin-boson
coupling~\eqref{spin_boson} seems to introduce a source of decoherence in the spin chain, thus hindering rather than
assisting the generation of stationary entanglement. In the following section, we will show that this na\"{i}ve
intuition is not
always valid, and explain the subtle mechanism that allows for generation of stationary entanglement in a particular
regime of the system parameters.

\subsection{Cooling or heating to an entangled steady state}
\label{cooling_spin_wave}

We consider an initial state of the system $\rho(0)=\ket{\psi_{0}}\bra{\psi_{0}}\otimes\rho_{\rm b}(0)$,
where  $\rho_{\rm b}(0)$ is an arbitrary state of the harmonic mode, and the uncorrelated spin state
$\ket{\psi_{0}}=\ket{{\uparrow_1\downarrow_2\downarrow_3\cdots\downarrow_N}}$ has a single spin excitation at 
the left edge
of the chain. In this section, we discuss how the system Liouvillian $\mathcal{L}_{\rm DSBC}$ acts on this state,
allowing for the onset of stationary entanglement. 

\subsubsection{\it Spin-chain spectrum} 
As a consequence of the XY Hamiltonian~\eqref{xy_model}, the initially localized spin excitation is exchanged between
different neighbors. In fact, the spin chain corresponds exactly to a tight-binding model where the excitation hops
along the sites of the underlying chain according to 
\begin{equation}
\label{xy_one_excitation}
H_{{\rm s},1}=\mathcal{P}_{\rm s}H_{\rm s}\mathcal{P}_{\rm s}=\sum_{i=1}^{N-1}J\ket{i}\bra{i+1}+\text{H.c.}.
\end{equation}
In this expression, $H_{{\rm s},1}$ results from the projection of the XY Hamiltonian onto the single-excitation
subspace $H_{{\rm s}}\to\mathcal{P}_{\rm s}H_{\rm s}\mathcal{P}_{\rm s}$,  where $\mathcal{P}_{\rm s}$ is the projector
onto an $N$-dimensional subspace spanned by the states $\ket{i}$, such that $i$ labels the position of the spin
excitation in the chain.  This tight-binding model gives rise to the following band structure (see
Fig.~\ref{fig_scheme}{\bf (b)})
\begin{equation}
\epsilon_n=2J\cos(q_{n}),\hspace{3ex}q_n=\frac{\pi n}{N+1},\hspace{1ex}n=1\cdots N,
\end{equation}
where each energy is associated to a ``spin-wave excitation'' 
\begin{equation}
\label{spin_wave}
\ket{q_n}=\mathcal{N}\sum_{i=1}^N\sin(q_ni)\ket{i},\hspace{3ex}\mathcal{N}=\left(\frac{2}{N+1}\right)^{\half}
\end{equation}
The goal of this section is to find the conditions to generate dissipatively one of these multipartite entangled states,
${\rm Tr}_a\{\ee^{\mathcal{L}_{\rm DSBC}t}\ket{\psi_0}\bra{\psi_0}\otimes\rho_{\rm b}(0)\}\to\ket{q_n}\bra{q_n}$.

\subsubsection{\it Spin-wave ladder} We consider a spatial modulation of the spin-boson coupling
strength~\eqref{spin_boson} in
the form
$g_i=g\cos(q_{\rm g}i)$, where $q_{\rm g}=\pi/(N+1)$. In the single-excitation subspace, the spin-boson coupling becomes
\begin{equation}
\label{projected_sb}
H_{{\rm sb},1}=\mathcal{P}_{\rm s}H_{\rm sb}\mathcal{P}_{\rm s}=\sum_{n=1}^{N-1}g(L_n^++L_n^-)(a+a^{\dagger}),
\end{equation}
where  the ladder operators $L_n^+=\ket{q_n}\bra{q_{n+1}}=(L_n^-)^{\dagger}$ are responsible for ascending or descending
along the ladder of spin-wave excitations (see Fig.~\ref{fig_scheme}{\bf (b)}). Accordingly, the spin-boson
coupling~\eqref{projected_sb} connects different spin-wave excitations, while simultaneously  creating or annihilating 
bosonic quanta. Moreover, since the harmonic mode dissipates irreversibly into a Markovian environment~\eqref{damping},
the ladder operators  $L_n^{\pm}$ introduce irreversible dynamics in the spin chain. In terms of these ladder
operators, the single-excitation spin-chain Hamiltonian~\eqref{xy_one_excitation} reads 
\begin{equation}
H_{\rm s,1}=\sum_{n=1}^{N-1}\epsilon_nL_{n}^{+}L_n^{-}\,,
\end{equation}
where for simplicity we reset the zero of energy at the bottom of the band. 
We show below how to generate  mesoscopic entangled
states by controlling the relative strengths of the dissipative processes that take the system up and down the {\it
spin-wave
ladder}.

\subsubsection{\it Irreversible dynamics in the spin chain}

 Since we are interested in the stationary properties of the DSBC, we
consider long times $t\gg\kappa^{-1}$, such that the bosonic mode has enough time to relax under the action of the
damping. Besides, for $|g|\ll\kappa$, this relaxation  is much faster than the process of energy exchange between
the boson and the spins. In this limit, the boson thermalizes individually, and the spin chain evolves on a much slower
timescale under such a bosonic background.  

To obtain the effect of the boson background on the slower spin dynamics~\cite{ad_elim}, we must  ``integrate out'' the
bosonic degrees of freedom from the spin-boson coupling~\eqref{spin_boson}. We start with a state of the form $\rho
(t) = \rho_{\rm b}^{\rm ss} \otimes\rho_{\rm s} (t)$, where $\rho_{\rm b}^{\rm ss}$ is the vacuum of the
harmonic mode, and fulfills $\mathcal{D}_{\rm b}\rho_{\rm b}^{\rm ss}=0$. We then expand the state at a later time
$t+\delta t$, with $\kappa\delta t\gg1$, in powers of the coupling constant $g$ between the spin system and the
oscillator, keeping up to second order in $g$. Tracing over the mode, we obtain:
\begin{multline}
 \hat{\rho_{\rm s}}(t+\delta t) \simeq \hat{\rho_{\rm s}} (t) + \int_{t}^{t+\delta t} {\rm d}t' \int_{t}^{t'} {\rm
d}t'' \,{\rm
Tr}_{\rm b}\Big\{\hat{\mathcal{L}}_{\rm sb} (t') e^{\mathcal{D}_{\rm b} (t'-t'')}\\ \hat{\mathcal{L}}_{\rm sb} (t'')
\rho_{\rm b}^{\rm ss} \otimes \hat{\rho}_{\rm s} (t) \Big\}
\end{multline}
where the ``hat'' indicates that we work in interaction picture with respect to $H_{\rm s,1}+H_{\rm b}$, and where we
have
introduced $\mathcal{L}_{\rm sb}(\bullet)=-\ii[H_{\rm sb,1},\bullet ]$. Using the explicit form of
$\mathcal{D}_{\rm b}$ and $\mathcal{L}_{\rm sb}$, after
some
algebra and one integral in $t''$ we find:
\begin{multline}
\label{first_integral}
\hat{\rho}_{\rm s}(t+\delta t)\simeq \hat{\rho}_{\rm s} (t) + \frac{|g|^2}{\kappa} \int_{t}^{t+\delta t} {\rm d}t'
\bigg[ \hat{J}_{{\rm
coll}}^{(2)}(t') \hat{\rho}_{\rm s} (t) \hat{J}_{{\rm coll}}^{(1)} (t') \\
-\hat{J}_{{\rm coll}}^{(1)}(t') \hat{J}_{{\rm coll}}^{(2)}(t') \hat{\rho}_{\rm s} (t) +\text{H.c.} \bigg].
\end{multline}
Here, we introduced the {\it collective jump operators}
\begin{equation}
\label{first_jump_operator}
J_{{\rm coll}}^{(1)}=\!\!\sum_{n=1}^{N-1}(L_n^++L_n^-),\hspace{2ex}J_{{\rm
coll}}^{(2)}=\!\!\sum_{n=1}^{N-1}(\xi_n^+L_n^++\xi_n^-L_n^-),\\
\end{equation}
with  $\xi_n^{\pm}=\kappa/[\kappa + \ii(\Delta_a\pm\Delta_n)]$, being $\Delta_n=\epsilon_{n}-\epsilon_{n+1}>0$ the
energy difference between two neighboring spin-wave excitations in the spin-wave ladder (see Fig.~\ref{fig_scheme}{\bf
(b)}).  

The integrand in Eq. (\ref{first_integral}) contains some terms that oscillate rapidly in time. We now perform the
remaining integral, assuming that the time interval $\delta t$ is short compared to the time scale
given by $\kappa/g^2$, but long enough such that $\delta t |\Delta_n-\Delta_{n'}|\gg 1~\forall \Delta_n\neq\Delta_{n'}$.
This condition restricts the values of $g$ for which this treatment
is valid. To perform the integral, we group the frequencies $\Delta_n$ in such a way that within each group the
frequencies are equal, and the difference between the frequencies in different groups is finite (we note that this is
not possible in the limit of
infinite sizes, where the energies become a continuum). It is worth noticing that the presence of degenerate frequencies
is typical in this model and makes the
grouping necessary. If one keeps only the dominant terms, which are the terms in the integrand that are
constant in time, back in Schr\"odinger picture one finds the following master equation governing the coarse-grained
evolution over the time scales of interest:
\begin{equation}
\label{effective_liouvillian}
\frac{{\rm d}{\rho}_{\rm s}}{{\rm d}t}=\mathcal{L}_{\rm s}(\rho_{\rm s})=-\ii[H_{\rm eff},\rho_{\rm
s}]+\mathcal{D}_{\rm s} (\rho_{\rm s}),
\end{equation}
with
\begin{multline}
H_{\rm eff} = H_{\rm s,1} + 2\pi \sum_\Delta \Bigg[\frac{\Delta_a-\Delta}{2\kappa} J_a(\Delta) J_\Delta^+ J_\Delta^-
\\
+ \frac{\Delta_a+\Delta}{2\kappa} J_a(-\Delta) J_\Delta^- J_\Delta^+ \Bigg]\,,
\end{multline}
and where
\begin{multline}
\label{complete_dissipator}
\mathcal{D}_{\rm s} \rho_{\rm s} = 2\pi \sum_\Delta \Bigg[ J_a(\Delta) \left( J_\Delta^- \rho_{\rm s} J_\Delta^+ -
\frac{1}{2} \{ J_\Delta^+J_\Delta^-, \rho_{\rm s}\} \right) \\
+ J_a(-\Delta) \left( J_\Delta^+ \rho_{\rm s} J_\Delta^- - \frac{1}{2} \{ J_\Delta^- J_\Delta^+, \rho_{\rm s}\} \right)
\Bigg]\,.
\end{multline}
In the expressions above, the sum over $\Delta$ runs over the different transition frequencies in the system,
$\{\,,\}$ denotes an anticommutator, and the Lindblad operators are defined as:
\beq
\label{jump_operator}
J_\Delta^+ = \sum_{n/\Delta_n=\Delta} L_n^+ = {J_\Delta^-}^\dagger
\eeq
where the sum is over all the values of $n$ such that $\Delta_n=\Delta$. The action of the different Lindblad
operators is weighted by the spectral density of the effective environment, 
\begin{equation}
\label{spectral_density}
J_{a}(\omega)=\frac{\kappa}{\pi}\frac{|g|^2}{[\kappa^2+(\omega-\Delta_a)^2]}.
\end{equation}
According to these expressions, the spins are subjected to a Lorentzian reservoir centered at the boson
detuning $\Delta_a$ with a width given by the boson damping rate $\kappa$. Therefore, the dissipation on the spins 
is not equal at all frequencies, but stronger at frequencies  matching that of the bosonic mode (i.e. they do
not see a totally flat environment, but a structured one~\cite{lorentzian}). 

As announced previously, the ladder operators $L_n^{\pm}$ are responsible for introducing the irreversible dynamics in
the spin chain. In particular, they determine the collective jump operators~\eqref{jump_operator}, where the adjective
collective emphasizes that they act over all the spins in the chain. However, we remark that these jump operators are a
sum of single-spin operators, as opposed to the multi-spin nature of some other engineered dissipation protocols
considered recently~\cite{dissipative_pure_state_engineering,
dissipative_qc,rydberg_dissipation,digital_dissipation_ions}. With this discussion, we show the third property of the
protocol outlined in the introduction: jump operators are composed of sums of {\it (iii) single-qubit operators}.  This
draws an analogy to the models of collective spontaneous emission~\cite{dicke}, but we will show that the special
dissipation mediated by the boson mode allows us to use them to prepare dissipatively the ground state of the spin
chain.

\begin{figure}

\centering
\includegraphics[width=1\columnwidth]{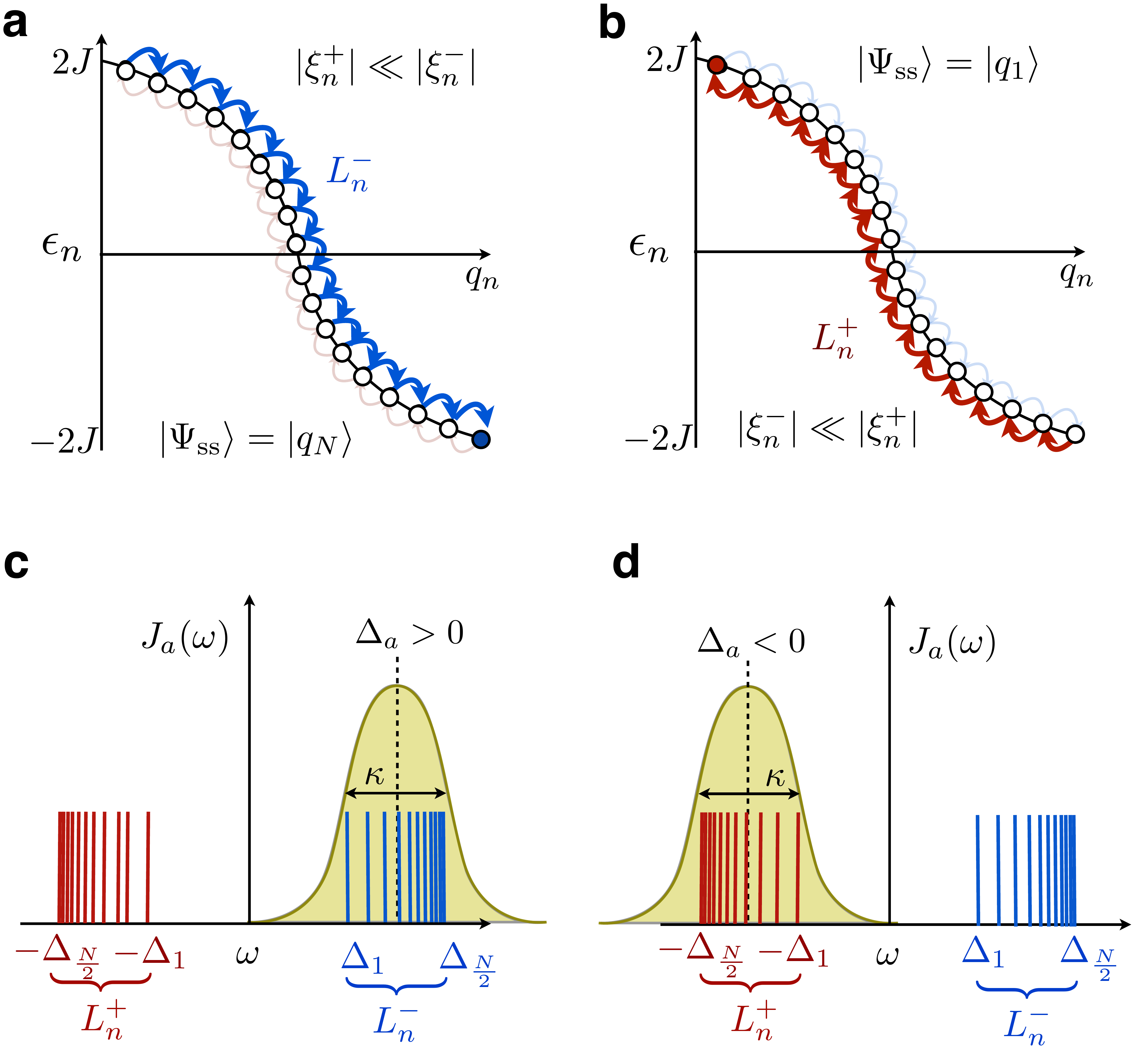}
\caption{ {\bf Tailoring the dissipative jump operators:} {\bf (a,c)}  In the regime of positive oscillator detuning
$\Delta_a\approx\Delta_n$, the jump operators that descend the ladder govern the dissipative dynamics (i.e. $r_n\ll1$).
This can be understood in terms of the effective Lorentzian spectral density $J_a$ of the environment seen by the  spin
chain, which
overlaps minimally with the processes that climb up the ladder. {\bf (b,d)}  In the regime of negative oscillator
detuning $\Delta_a\approx-\Delta_n$, the jump operators that climb up the ladder dominate the dissipative dynamics (i.e.
$r_n\gg1$), since in this case the spectral density of the environment seen by the spin
chain overlaps minimally with the processes that descend the ladder.}
\label{fig_2}
\end{figure}

\subsubsection{\it Stationary entanglement}

 We are searching for the steady state $\rho_{\rm s}^{\rm ss}$  of Eq.~\eqref{effective_liouvillian}, $\mathcal{L}_{\rm
s}(\rho_{\rm s}^{\rm ss})=0$, with three properties: being pure,  displaying multipartite entanglement, and being
unique. A pure steady state $\rho_{\rm s}^{\rm ss}=\ket{\Psi_{\rm ss}}\bra{\Psi_{\rm ss}}$ necessarily belongs to a
Hamiltonian eigenspace~\cite{dissipative_pure_state_engineering}, which in our case is given by the spin-wave
excitations $\ket{\Psi_{\rm ss}}\in\{\ket{q_n},n=1\cdots N\}$. Additionally, the steady state should also
belong to the kernel of the dissipator, 
$\mathcal{D}_{\rm s}(\ket{\Psi_{\rm ss}}\bra{\Psi_{\rm ss}})=0$. By inspection of the jump operators, we identify two
regimes where this can happen, which depend on  the boson detuning with respect to the spin-wave energy difference:
\begin{equation}
\label{conditions_down}
\kappa\ll+\Delta_a\approx\Delta_n\Rightarrow |\Delta_a-\Delta_n|\ll|\Delta_a+\Delta_n|,\hspace{1ex}r_n\to 0,\hspace{1ex}
\end{equation}
\begin{equation}
\label{conditions_up}
\kappa\ll-\Delta_a\approx\Delta_n\Rightarrow|\Delta_a+\Delta_n|\ll|\Delta_a-\Delta_n|,\hspace{1ex}r_n\to \infty.
\end{equation}
For the regime of positive detunings, we can then approximate the evolution using that
$J_a(-\Delta_n)$ is negligibly small for all $n$, while all the
$J_a(\Delta_n)$ are non-negligible. In that case the non-unitary part of the evolution is 
\beq
\mathcal{D}_{\rm s} \rho_{\rm s} \simeq 2\pi \sum_\Delta J_a(\Delta) \left( J_\Delta^- \rho_{\rm s} J_\Delta^+ -
\frac{1}{2}
\{ J_\Delta^+ J_\Delta^-, \rho_{\rm s}\} \right)\,,
\eeq
taking the system to the lowest state in the manifold of spin waves (the Hamiltonian part does not induce transitions
between spin waves). Thus, the unique pure steady state corresponds to the lowest spin wave  $\ket{\Psi_{\rm
ss}}=\ket{q_N}$.
Conversely, for negative detunings, the steady state corresponds
to the highest spin-wave $\ket{\Psi_{\rm ss}}=\ket{q_1}$.
Following~\cite{dissipative_pure_state_engineering}, it can be shown that in these limits the particular form of
the Lindblad
operators does not allow for mixed steady states. We thus get the desired result: a unique pure state displaying
stationary multipartite entanglement.

In Fig.~\ref{fig_2}, we represent schematically the two regimes of interest. When 
conditions~\eqref{conditions_down} are fulfilled, for $\kappa\ll\Delta_a\approx\Delta_n$, the
dissipative processes that go down the spin-wave ladder dominate since $|\xi_n^+|\ll|\xi_n^-|$. Effectively, the
dissipation induced by the damped bosonic mode ``cools'' the spin state to the lowest-energy spin wave, ${\rm
Tr}_{\rm b}\{\ee^{\mathcal{L}_{\rm DSBC}t}\ket{\psi_0}\bra{\psi_0}\otimes\rho_{\rm b}(0)\}\to\ket{q_N}\bra{q_N}$ (see
Fig.~\ref{fig_2}{\bf (a)}). The bath spectral
density peaks at the frequencies corresponding to
the dissipative processes descending the spin-wave ladder (see Fig.~\ref{fig_2}{\bf (c)}) and therefore, the bath
absorbs
very efficiently the energy dissipated by the spin chain during its relaxation to the spin wave
$\ket{\Psi_{\rm ss}}=\ket{q_N}$. Conversely, when  the system parameters fulfill conditions~\eqref{conditions_up},
for negative detunings $\kappa\ll-\Delta_a\approx\Delta_n$, it is the
dissipative processes that climb up the ladder which dominate, $|\xi_n^-|\ll|\xi_n^+|$. In this limit, the bosonic mode
drives the spin state to the highest-energy spin wave, ${\rm Tr}_{\rm b}\{\ee^{\mathcal{L}_{\rm
DSBC}t}\ket{\psi_0}\bra{\psi_0}\otimes\rho_{\rm b}(0)\}\to\ket{q_1}\bra{q_1}$ (see Fig.~\ref{fig_2}{\bf (b)}).
The
bath spectral density  is maximal for the ascending dissipative processes (see Fig.~\ref{fig_2}{\bf (d)}), and the bath
provides the required energy to climb up the spin-wave ladder. We may thus conclude that in both regimes, the structured
reservoir singles
out a unique spin wave as the steady state of the chain, assisting the generation of stationary multipartite
entanglement. 

In the light of these results, we can also understand the regime $\kappa\gg J$. In this limit, the reservoir spectral
density~\eqref{spectral_density} becomes essentially flat, and both up/down processes contribute equally
$\xi_n^+\approx\xi_n^-\approx 1$. It is straightforward to see that if $J_a(\omega)$ is constant, the dissipator in
Eq. (\ref{complete_dissipator}) satisfies
$\mathcal{D}_{\rm s} (\mathbb{I}) = 0$, and the
totally mixed state $\rho_{\rm s}^{\rm ss}\propto\mathbb{I}$ becomes a steady state. This regime corresponds to the
na\"{i}ve argument of section~\ref{model}, following which one expects
that the dephasing-like term~\eqref{spin_boson} can only decohere the spin chain. It is now clear that there are other
regimes,~\eqref{conditions_down}-\eqref{conditions_up}, where we can profit from a structured environment to assist the
generation of stationary entanglement.

We now comment on the time required to reach the aforementioned steady states.
In the regime where the effective Liouvillian~\eqref{effective_liouvillian} was derived, it is given by $t_{\rm f}\sim
(|g|^2/\kappa)^{-1}$, where $|g|\ll\kappa\ll J$. For a
particular experimental setup, where the spin couplings cannot reach arbitrarily large values, this preparation time
may turn out to be too long for practical purposes. We emphasize, however, that  the same dissipative preparation of
entangled states can be obtained in a non-Markovian regime where $|g|\sim\kappa\ll J$, or $\kappa\ll|g|\ll J$. Since
simple master equations cannot be derived analytically for this regime, we shall explore it numerically.

\subsubsection{\it Dissipative preparation of W-like states}

 Let us note that although the spin-wave excitations~\eqref{spin_wave} are genuinely multipartite entangled, the weights
for the excitation of each of the spins in the system are different. A small modification of
the scheme also allows for
the dissipative generation of $N$-partite W-like states, where the W-state is defined as
\beq
\ket{W}=\frac{1}{\sqrt{N}}(\ket{{\uparrow\downarrow\downarrow\cdots\downarrow}}+\ket{{
\downarrow\uparrow\downarrow\cdots\downarrow}}+\cdots\ket{{\downarrow\downarrow\downarrow\cdots\uparrow}}). 
\eeq
Let us leave for a moment the constraint to consider purely $\it global$ couplings, and assume that it is possible to
add external transverse fields that act locally at the two edges of the chain
\begin{equation}
\label{edge_transverse_fields}
H_{\rm s}=J\left(\sum_{i=1}^{N-1}(\sigma_i^+\sigma_{i+1}^-+\text{H.c.})-\frac{1}{2}\sigma_1^z - \frac{1}{2}\sigma_N^z
\right).
\end{equation}
In the subspace with one spin excitation, the ground state of this Hamiltonian is of the form
\beq
\ket{\tilde{q}_N} = \frac{1}{\sqrt{N}} \sum_{i=1}^N (-1)^i \ket{i},
\eeq
which is locally equivalent to the above $N$-partite W-state (we note that the actual W-state can be
obtained if the spin-spin coupling is ferromagnetic instead of antiferromagnetic).
As opposed to the ground state of the spin system considered so far, the W-state has the property that it is fully
symmetric under particle interchanges. In the state $\ket{\tilde{q}_N}$, the excitation is equally distributed over all
the sites, and each of the particles is equally entangled with the rest. This state can be prepared
with the same method described before, at the price of introducing individual addressability in the trapped-ion
proposal of Sec.~\ref{sec:ions}.

\subsubsection{\it Numerical analysis of the protocol}

\begin{figure}
 \includegraphics[width=0.8\columnwidth]{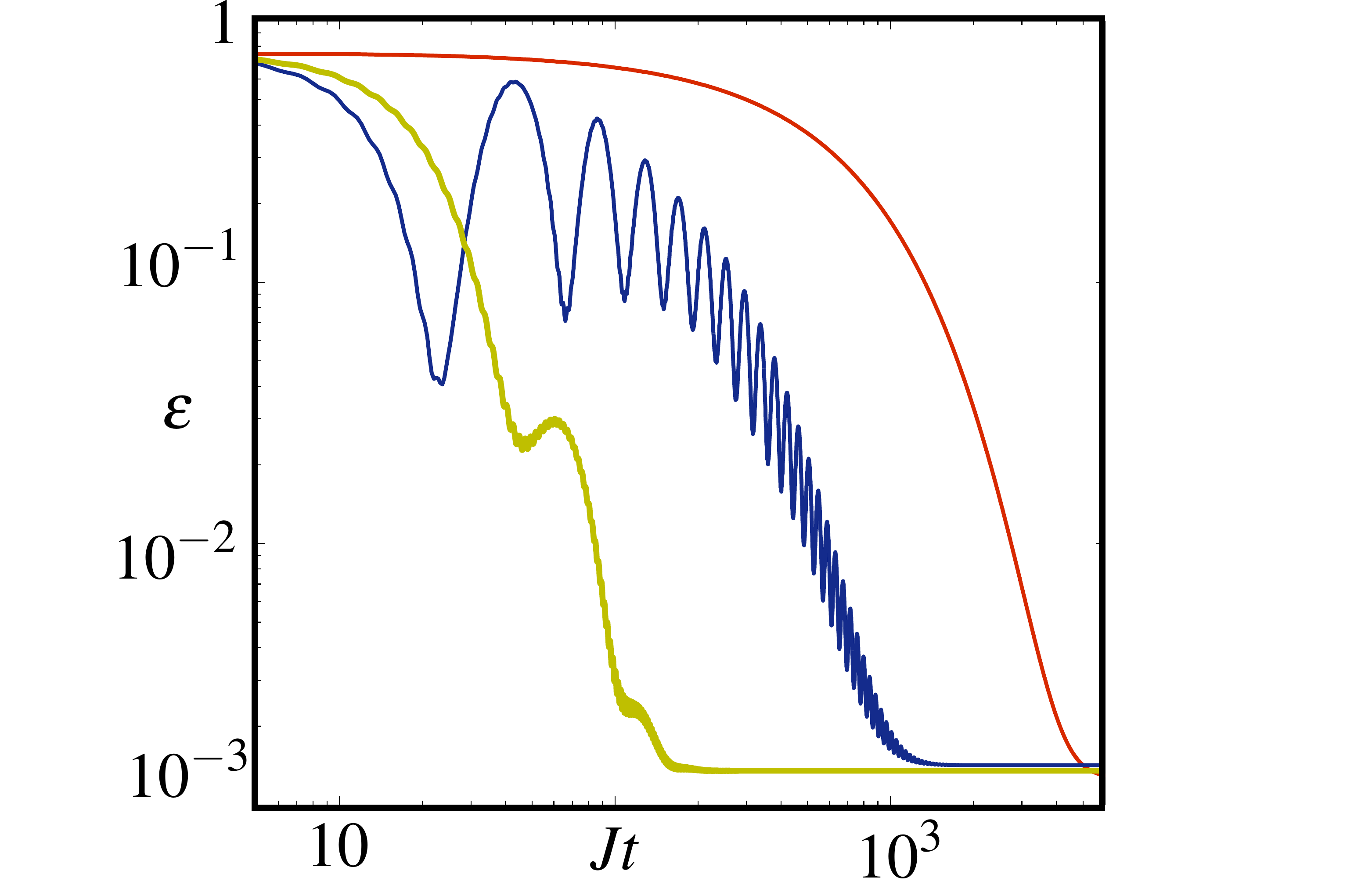}
 \caption{\label{fig:fidelity_dynamics} {\bf Approach to the asymptotic state:} Time evolution of the error $\varepsilon
=
1-\mathcal{F}$, where $\mathcal F$ is the fidelity with the inhomogeneous spin-wave excitation
$\ket{q_3}=(\ket{{\uparrow\downarrow\downarrow}}-\sqrt{2}\ket{{\downarrow\uparrow\downarrow}}+\ket{{
\downarrow\downarrow\uparrow}})/2$ for a chain of three spins. The curves correspond to the resonant case
$\Delta=\sqrt{2}J$, while the other
parameters are $g=0.01$, $\kappa=0.1$ (red), $g=0.075$, $\kappa=0.0075$ (blue), and $g=\kappa=0.06$ (green).
For these numerical calculations, the oscillator was truncated to three levels, with trace preserved to order
$10^{-12}$.} 
\end{figure}

\begin{figure*}
 \includegraphics[width=1.3\columnwidth]{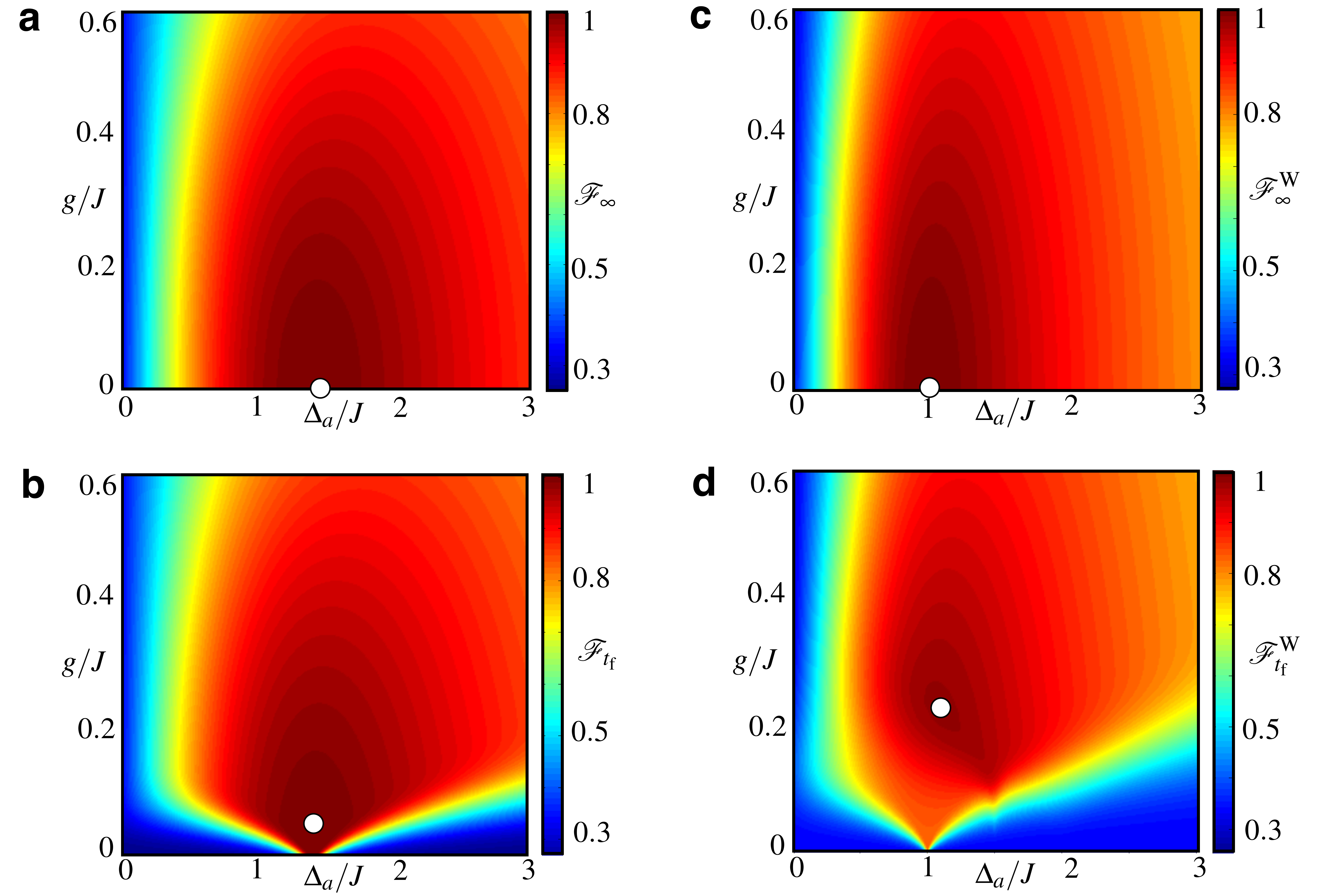}\\
 \caption{\label{fig:fidelity_N3} {\bf Stationary tripartite entangled state:} {\bf (a)} Contour plot of the asymptotic
fidelity $\mathcal{F}_{\infty}$ for generating the inhomogeneous spin-wave excitations
$\ket{q_3}=(\ket{{\uparrow\downarrow\downarrow}}-\sqrt{2}\ket{{\downarrow\uparrow\downarrow}}+\ket{{
\downarrow\downarrow\uparrow}})/2,
$ as a function of the bosonic detuning $\Delta_a$ (horizontal axis), and the spin-boson coupling
$g=\kappa$ (vertical axis), for a chain of $N=3$ sites (the optimal condition~\eqref{conditions_down} is met
for
$\Delta_a= \sqrt{2}J$, and $g\ll J$). {\bf (b)} The same as in {\bf (a)} but setting a finite time $t_{\rm f} =
10^3/J$. 
{\bf (c)} Contour plot of the asymptotic fidelity $\mathcal{F}_{\infty}$ for generating the W-like states
$\ket{\tilde{q}_3}=(\ket{{\uparrow\downarrow\downarrow}}-\ket{{\downarrow\uparrow\downarrow}}+\ket{{
\downarrow\downarrow\uparrow}})/\sqrt{3},
$ as a function of the bosonic detuning $\Delta_a$ (horizontal axis), and the spin-boson coupling
$g=\kappa$ (vertical axis), for a chain of $N=3$ sites (the optimal condition~\eqref{conditions_down} is met
for
$\Delta_a= J$, and $g\ll J$). {\bf (d)} The same as in {\bf (c)} but setting a finite time $t_{\rm f} = 10^3/J$. 
For these numerical 
calculations, the oscillator was truncated to three levels, the trace is preserved to $10^{-6}$ and the maximum
fidelity is equal to 1 within this error. In each plot, the highest fidelity is indicated by a white dot.} 
\end{figure*}

 In this section, we analyze numerically the validity of our previous analytical derivations  by integrating directly
the master equation of the damped spin-boson chain~\eqref{dsbc}. We obtain the fidelity with the desired
stationary entangled state
\begin{equation}
\mathcal{F}_{\ket{\Psi_{\rm target}}}=|\bra{\Psi_{\rm target}}{\rm Tr}_{\rm b}\{\ee^{\mathcal{L}_{\rm
DSBC}t}\ket{\psi_0}\bra{\psi_0}\otimes\rho_{\rm b}(0)\}\ket{\Psi_{\rm target}}|,
\end{equation}
where $\ket{\psi_{0}}=\ket{{\uparrow_1\downarrow_2\downarrow_3\cdots\downarrow_N}}$, we consider that the boson mode is
initially in the vacuum state $\rho_{\rm b}(0)=\ket{0}\bra{0}$, and ${\ket{\Psi_{\rm target}}}$ is the particular
entangled state that the protocol targets. We test the  robustness of this fidelity against a non-optimal choice of the
system parameters~\eqref{conditions_down}-\eqref{conditions_up}, a limited protocol time $t_{\rm f}$, and an increasing
particle number $N$. 

In Figure \ref{fig:fidelity_dynamics}, we consider the time evolution of the fidelity for the preparation of the
lowest-energy spin-wave excitation of an $N=3$ spin chain
 \begin{equation}
 \label{spin_wave_3}
 \ket{q_3}=\frac{1}{2}\ket{{\uparrow\downarrow\downarrow}}-\frac{1}{\sqrt{2}}\ket{{\downarrow\uparrow\downarrow}}+\frac{
1}{2}\ket{{\downarrow\downarrow\uparrow}}.
 \end{equation}
For these calculations, we set the oscillator detuning $\Delta_a$ to the value $\Delta_a^{\star}=\sqrt{2}J$ which is
resonant with the spin-wave transitions, and vary the ratio
between the spin-boson coupling strength $g$ and the damping rate $\kappa$. The plot indicates that, for similar values
of the final fidelity, the convergence is much faster when $g$ and $\kappa$ are of the same order. This implies that for
practical purposes it is most convenient to work in the deeply non-Markovian regime. The optimal
ratio $g/\kappa$, however, depends in general on the number of spins and the value of $\Delta_a$.

It is important to note that the resonance condition can be relaxed to some extent. In
Figure~\ref{fig:fidelity_N3} we analyze the fidelity as a function of the detuning $\Delta_a$ and of the
coupling $g$, setting $\kappa=g$. From our previous analysis, we expect the fidelity to be
maximal when $\Delta_a=\Delta_a^{\star}$, and when $g, \kappa$ approach zero to fulfill the requirement
$g,\kappa\ll J$. This behavior is totally
captured by the numerical results in Fig.~\ref{fig:fidelity_N3}{\bf (a)}, displaying the asymptotic fidelity. The
maximum fidelity, highlighted with a white dot, is indeed found in this limit and is equal to 1 within the numerical
errors. We remark that, even when the
above conditions are only partially fulfilled, the dynamics can provide the desired state with fidelities well above
$90\%$. Therefore, we can claim that our dissipative protocol is considerably robust against parameter imperfections. 
 
Nevertheless, the time required to approach the desired target state might be prohibitively long. Therefore, we 
also characterize the fidelity for a fixed protocol duration. In Fig.~\ref{fig:fidelity_N3}{\bf (b)}, we show the value
of the fidelity at a finite fixed time $t_{\rm f} = 10^3/J$. As can be seen in this figure, the optimal choice for the
oscillator detuning is still $\Delta_a^{\star}=\sqrt{2}J$,  while the optimal value for $g=\kappa$ has moved away from
zero. In fact, the best choice now results from the interplay between
the condition $g,\kappa\ll\Delta_a,J$, and the required convergence speed. We emphasize that it is still
possible to find parameters such that the achieved fidelities remain very high, $\mathcal{F}_{t_{\rm
f}}^{\star}=0.9999$.

We now address the dissipative generation of W-like states by adding boundary transverse
fields~\eqref{edge_transverse_fields}. For the $N=3$ spin chain under consideration, the target state is 
 \begin{equation}
 \ket{\tilde{q}_3}=\frac{1}{\sqrt{3}}(\ket{{\uparrow\downarrow\downarrow}}-\ket{{
\downarrow\uparrow\downarrow}}+\ket{{\downarrow\downarrow\uparrow}}).
 \end{equation}
In Fig.~\ref{fig:fidelity_N3}{\bf (c)-(d)}, we show the asymptotic and finite-time fidelities for the dissipative
generation of such a tripartite entangled state with homogeneously-distributed excitations. We observe  high fidelities
comparable to the inhomogeneous spin-wave state~\eqref{spin_wave}, and moreover, an analogous robustness with respect to
a non-optimal choice of the system parameters. 
Let us note that, for $N=3$, there are two different transitions in the spectrum of these W-type spin-waves: one is
resonant at a frequency of $\tilde{\Delta}_2=J$, and the remaining one occurs at $\tilde{\Delta}_1=2J$. Therefore, it is
impossible to set an oscillator detuning at resonance with both transitions simultaneously, and thus the
conditions~\eqref{conditions_down} are not completely fulfilled. This explains the fact that the W-type fidelities
achieved for
finite times are lower than those corresponding to the inhomogeneous spin waves in Fig.~\ref{fig:fidelity_N3}{\bf (b)},
where the two transitions have the same frequency $\Delta_1=\Delta_2=\sqrt{2}J$.

In this section, we have analyzed numerically the validity of the scheme for the generation of the simplest case of
multipartite entanglement, namely tripartite entangled states. A question that should be carefully addressed is
whether the same scheme can be used to generate $N$-partite entangled states, and how large can the attained fidelities
be as $N$ is increased. This is the topic of the following section.

\subsubsection{\it Mesoscopic spin chains}

\begin{figure}
 \includegraphics[width=.9\columnwidth]{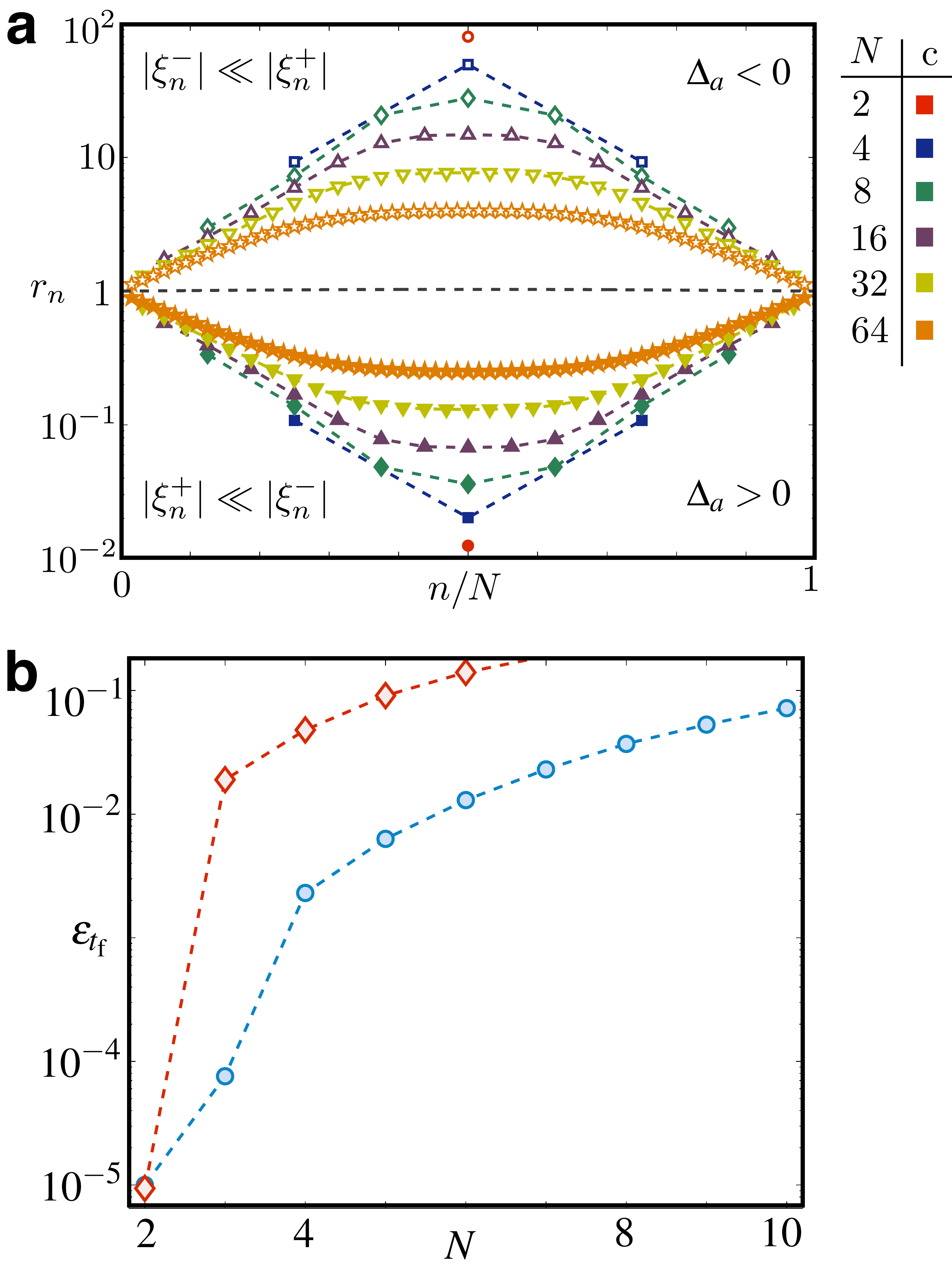}
 \caption{\label{fig:Nsites} {\bf The dissipative protocol in mesoscopic chains:} {\bf (a)} Ratio
$r_n=|\xi^{+}_n|/|\xi^{-}_n|$ of the relative strength of the dissipative processes climbing up and going down the
ladder as a function of quasimomentum index $n$ for different number of spins $N$. The values $r_n>1$ correspond to
negative
oscillator detunings $\Delta_a=-\Delta_{N/2}<0$ which select the processes up the ladder, whereas $r_n<1$ corresponds to
positive oscillator detunings $\Delta_a=\Delta_{N/2}>0$ selecting the cooling processes down the ladder. {\bf (b)}
 Error $\epsilon_{t_{\rm f}}=1-\mathcal{F}_{t_{\rm f}}$ for the dissipative generation of the target state in a finite
time $t_{\rm f}=10^3/J$ as a function of the number of spins $N$ in the chain. The blue circles represent the
errors corresponding to  the inhomogeneous spin waves $\ket{q_N}\propto
\sin(q_N)\ket{{\uparrow\downarrow\downarrow\cdots\downarrow}}+\sin(2q_N)\ket{{
\downarrow\uparrow\downarrow\cdots\downarrow}}+\cdots+\sin(Nq_N)\ket{{\downarrow\downarrow\downarrow\cdots\uparrow}}$.
The  red diamonds represent the error for the W-like states $\ket{\tilde{q}_N}\propto
\ket{{\uparrow\downarrow\downarrow\cdots\downarrow}}-\ket{{\downarrow\uparrow\downarrow\cdots\downarrow}}+\cdots+(-1)^{
N+1}\ket{{\downarrow\downarrow\downarrow\cdots\uparrow}}$. For the numerical calculation, the Hilbert space of the
oscillator was truncated to four levels.} 
\end{figure}

We start by assessing the fulfillment of the necessary conditions~\eqref{conditions_down}-\eqref{conditions_up}, which
rely on the
energetic difference between the dissipative processes that climb up/down the spin-wave ladder, as the number of spins
$N$ is increased. For very large spin
chains $N\to\infty$, the energy differences between neighboring spin waves $|\Delta_n|\lesssim 2\pi J/N\to0$. This
implies that the energetic argument selecting only processes going up or down the ladder can no longer hold. Indeed,
$\lim_{N\to\infty}r_n=\lim_{N\to\infty}|\xi^{+}_n|/|\xi^{-}_n|=1$, so that 
the protocol ceases to be operative in the thermodynamic limit of infinitely long chains.
 However, for mesoscopic spin chains, the ratio $r_n$ can be controlled to an acceptable degree. 
In Fig.~\ref{fig:Nsites}{\bf (a)}, we show that for positive detunings and $N\sim 10$, $r_n\approx0.1$ for the most of
the spin-wave
excitations, whereas slightly higher values are attained for the extremal spin excitations (where $|\Delta_n|\propto
J/N^2$). For ratios $r_n$ on this order, we expect  that  the dissipative preparation of entangled states still works
with acceptable fidelities.

To be more specific, we note that the presence of different transition frequencies $\Delta_n$ is generic for $N\geq4$,
which represents an
obstacle for the efficiency of the procedure. As the number of spins is increased, the convergence also becomes slower
because of the larger number of steps down/up the ladder towards the target state, and the lower transition frequencies
which require a decrease in $g$ and $\kappa$. 
In Fig. \ref{fig:Nsites}{\bf (b)}, we show numerical results for the dependence of the protocol error $\epsilon_{t_{\rm
f}}$
achieved for a finite time $t_{\rm f}=10^3/J$ as a function of the number of sites and optimizing the values of $g$,
$\kappa$ and $\Delta_a$.
We observe that the errors obtained for the dissipative state preparation of the inhomogeneous
spin wave $\ket{q_N}$ are below 10\% for chains of up to ten sites. In comparison, the
creation of the W-like states is worse, and such high fidelities can only be achieved for short chains of up to five sites.

\subsection{Preparation of the ground state of the isotropic XY chain in a transverse field}
\label{spin_gs_cooling}

So far, our analysis has been restricted to an $N$-dimensional subspace of the spin chain, since the initial spin state
$\ket{\psi_{0}}=\ket{{\uparrow_1\downarrow_2\downarrow_3\cdots\downarrow_N}}$ contains a single excitation, and the
number of spin excitations is preserved by the complete DSBC Liouvillian~\eqref{dsbc}. In the following we explore the
full $2^N$-dimensional Hilbert space of the spin chain.

\subsubsection{\it Jordan-Wigner jump operators}

 In order to treat the full Hilbert space of the spins, we fermionize the spin chain via the so-called Jordan-Wigner
transformation~\cite{jordan_wigner}, namely
\begin{equation}
\sigma_i^z=2c_i^{\dagger}c_i-1,\hspace{2ex}\sigma_i^+=c_i^{\dagger}\ee^{\ii\pi\sum_{j<i}
c_j^{\dagger}c_j}=(\sigma_i^-)^{\dagger},
\end{equation}
where $c_i^{\dagger},c_i$ are fermionic creation-annihilation operators. The spin-chain~\eqref{xy_model} and
spin-boson~\eqref{spin_boson} Hamiltonians can be expressed in terms of the Jordan-Wigner fermions as follows
\begin{equation}
H_{\rm s}=\sum_{i=1}^{N-1}Jc_i^{\dagger}c_{i+1}^{\phantom{\dagger}}+\text{H.c.},\hspace{2ex}H_{\rm
sb}=\sum_{i=1}^n2g_ic_i^{\dagger}c_i^{\phantom{\dagger}}(a+a^{\dagger}),
\end{equation}
where we recall that the spin-boson couplings are $g_i=g\cos(q_{\rm g}i)$ with $q_{\rm g}=\pi/(N+1)$. The next step is
to express these Hamiltonian terms in the spin-wave basis introduced in Eq.~\eqref{spin_wave}, which leads to the
following expressions
\begin{equation}
H_{\rm s}=\sum_{n=1}^{N}\epsilon_nc_{q_n}^{\dagger}c_{q_n}^{\phantom{\dagger}},\hspace{2ex} H_{\rm
sb}=\sum_{n=1}^{N-1}gc_{q_n}^{\dagger}c_{q_{n+1}}^{\phantom{\dagger}}\!(a+a^{\dagger})+\text{H.c.},
\end{equation}
where the fermionic operators in momentum space are $c_{q_n}=\mathcal{N}\sum_i\sin(q_ni)c_i$.
In combination with the dissipative part given by Eq.~\eqref{damping}, the spin-boson system is analogous to a {\it
damped
single-mode Holstein model}, a dissipative version of the familiar Holstein model describing electron-phonon
interactions~\cite{holstein}.

We can obtain the same formal expressions as in the single-excitation problem by rewriting the
ladder operators in second-quantized form
\begin{equation}
\label{JW_ladder}
L_n^{+}=\ket{q_n}\bra{q_{n+1}}=(L_n^{-})^{\dagger}\hspace{1ex}\to\hspace{1ex}
{L}^{+}_{{\rm f}, n} =c^{\dagger}_{q_n}c^{\phantom{\dagger}}_{q_{n+1}}=(L_{{\rm f}, n}^{-})^{\dagger}.
\end{equation}
Accordingly, in the regime where the boson degrees of freedom can be integrated out, we obtain a purely fermionic master
equation which coincides with Eqs.~\eqref{effective_liouvillian}-\eqref{complete_dissipator}, but with the collective
jump operators~\eqref{jump_operator} now expressed in terms of the Jordan-Wigner ladder operators
\begin{equation}
\label{jump_operator_fermions}
J_{{\rm f}, \Delta}^+ = \sum_{n/\Delta_n=\Delta} L_{{\rm f}, n}^+ = J_{{\rm f}, \Delta}^{-\dagger}.
\end{equation}
As a result, the dissipative dynamics restricted to the single-excitation sector can be generalized directly to the
full Hilbert space with arbitrary numbers of spin excitations.

\subsubsection{\it Effective ground-state cooling}

\begin{figure}
\centering
\includegraphics[width=1\columnwidth]{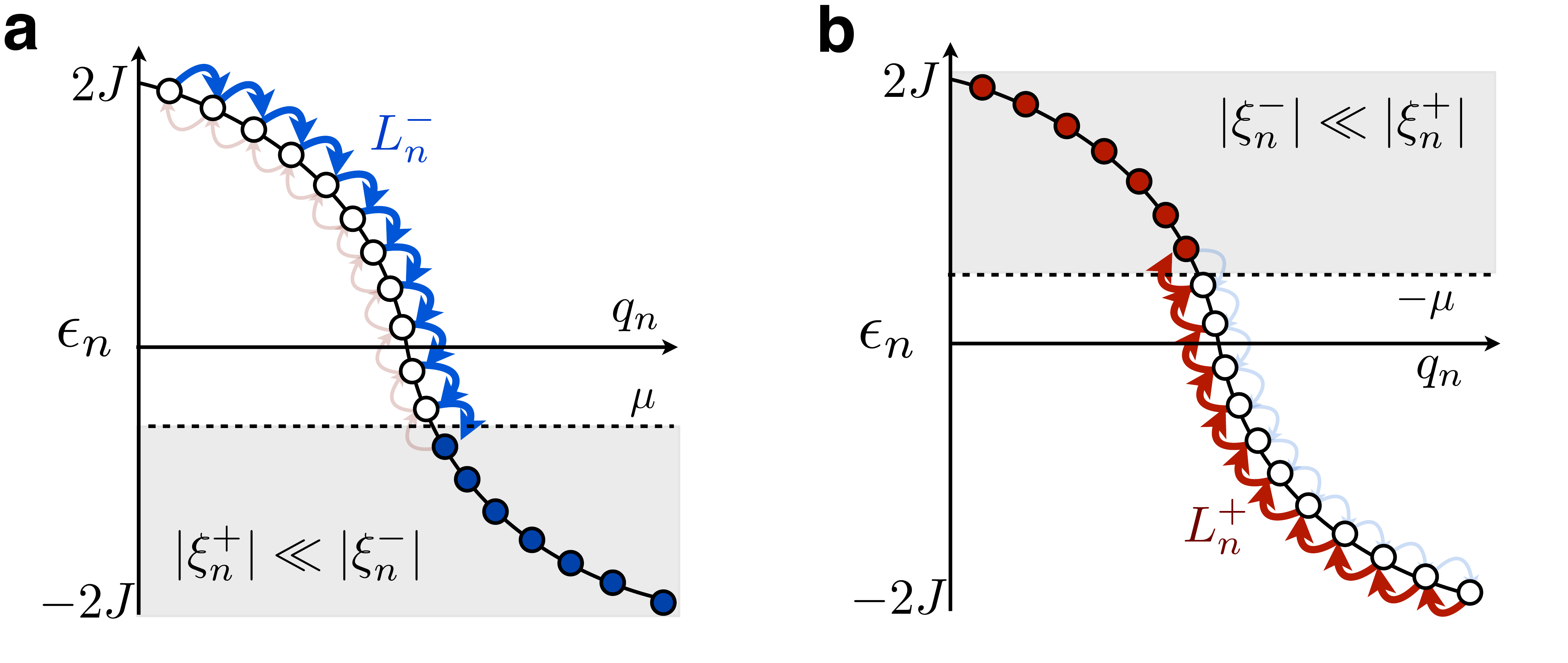}
\caption{ {\bf Cooling to the ground state of the spin chain:} {\bf (a)} Schematic representation of  the dissipative
processes that cool each of the $n_{\rm s}$ spin excitations to the lowest non-occupied state. The stationary state
corresponds to the ground state of the isotropic XY Hamiltonian, with the particular filling 
determined by an additional transverse field acting as a chemical potential. {\bf (b)} The same as in {\bf (a)} but
selecting the processes that climb up the ladder. The stationary state would correspond to the zero-temperature
state of the same XY model in a transverse field, but with reversed coupling strengths. }
\label{fig_6}
\end{figure}

 Let us now consider an initial state with an arbitrary number of spin excitations $n_{\rm s}\leq N$ distributed along
the chain
 \begin{equation}
 \ket{\psi_{0}}=\ket{{\uparrow_1\uparrow_2\cdots\uparrow_{n_{\rm s}}\downarrow_{n_{\rm s}+1}\downarrow_{n_{\rm
s}+2}\cdots\downarrow_N}}.
 \end{equation} 
 We would like to determine the steady-state of the spin-boson system if the conditions~\eqref{conditions_down} are
fulfilled. In this limit, the Lindblad operators are only of the form
\begin{equation}
J_{{\rm f}, \Delta}^- = \sum_{n/\Delta_n=\Delta} c^{\dagger}_{q_{n+1}} c^{\phantom{\dagger}}_{q_{n}}.
\end{equation}
Such a jump operator has the effect of lowering the energy of the fermionic quasiparticles. Due to the Pauli exclusion
principle, the initial $n_{\rm s}$ excitations cannot all occupy the lowest energy level,
with quasimomentum $q_N$. Instead, the stationary state must be of the following form
\begin{equation}
\label{gs_cooling}
\rho_{{\rm s}}^{{\rm ss}}=\ket{{\rm G_s}}\bra{{\rm G_{s}}},\hspace{2ex}\ket{{\rm G_{\rm s}}}=c_{q_{N-n_{\rm
s}+1}}^{\dagger}\cdots c_{q_{N-1}}^{\dagger}c_{q_{N}}^{\dagger}\ket{\rm vac}.
\end{equation} 
This is precisely the ground state of the original isotropic XY model~\eqref{xy_model}, if supplemented
with a homogeneous
transverse field $H_{\rm s}\to H(J,\mu)= H_{\rm s}-(\mu/2)\sum_i\sigma_i^z$. In this Hamiltonian, the
transverse field  plays the role of an effective chemical potential $\mu=\half(\epsilon_{N-(n_{\rm
s}-1)}+\epsilon_{N-n_{\rm s}})$, which is determined by the initial number of spin excitations. Due to the to the
dissipative process, the excitations are distributed in the lowest available single-particle states  (see
Fig.~\ref{fig_6}{\bf (a)}). This means, for each number of spins up in the initial state prepared, there is a value of
the transverse field $\mu$ such that the asymptotic state corresponds to the ground state of $H(J,\mu)$.
Conversely, a given choice of $J, \mu$ determines the number of spins that should be up in the initial state
so that the dissipative dynamics take the system into the desired ground state.

We could also explore the steady state when the conditions~\eqref{conditions_up} are fulfilled. In this limit, the
only Lindblad operators are of the form:
\begin{equation}
J_{{\rm f}, \Delta}^+ = \sum_{n/\Delta_n=\Delta} c^{\dagger}_{q_{n}}c^{\phantom{\dagger}}_{q_{n+1}}\,,
\end{equation}
and pump all the excitations to the highest-energy available single-particle states
\begin{equation}
\rho_{{\rm s}}^{{\rm ss}}=\ket{{\rm \tilde{G}_s}}\bra{{\rm \tilde{G}_{s}}},\hspace{2ex}\ket{{\rm \tilde{G}_{\rm
s}}}=c_{q_{n_{\rm s}}}^{\dagger}\cdots c_{q_{2}}^{\dagger}c_{q_{1}}^{\dagger}\ket{\rm vac}\,,
\end{equation} 
corresponding to the ground-state of the XY Hamiltonian $H(-J,-\mu)$ with a
ferromagnetic spin-spin coupling, and an inverted chemical potential (see Fig.~\ref{fig_6}{\bf (b)}).

\subsubsection{\it Numerical results}

\begin{figure}
\centering
 \includegraphics[width=0.9\columnwidth]{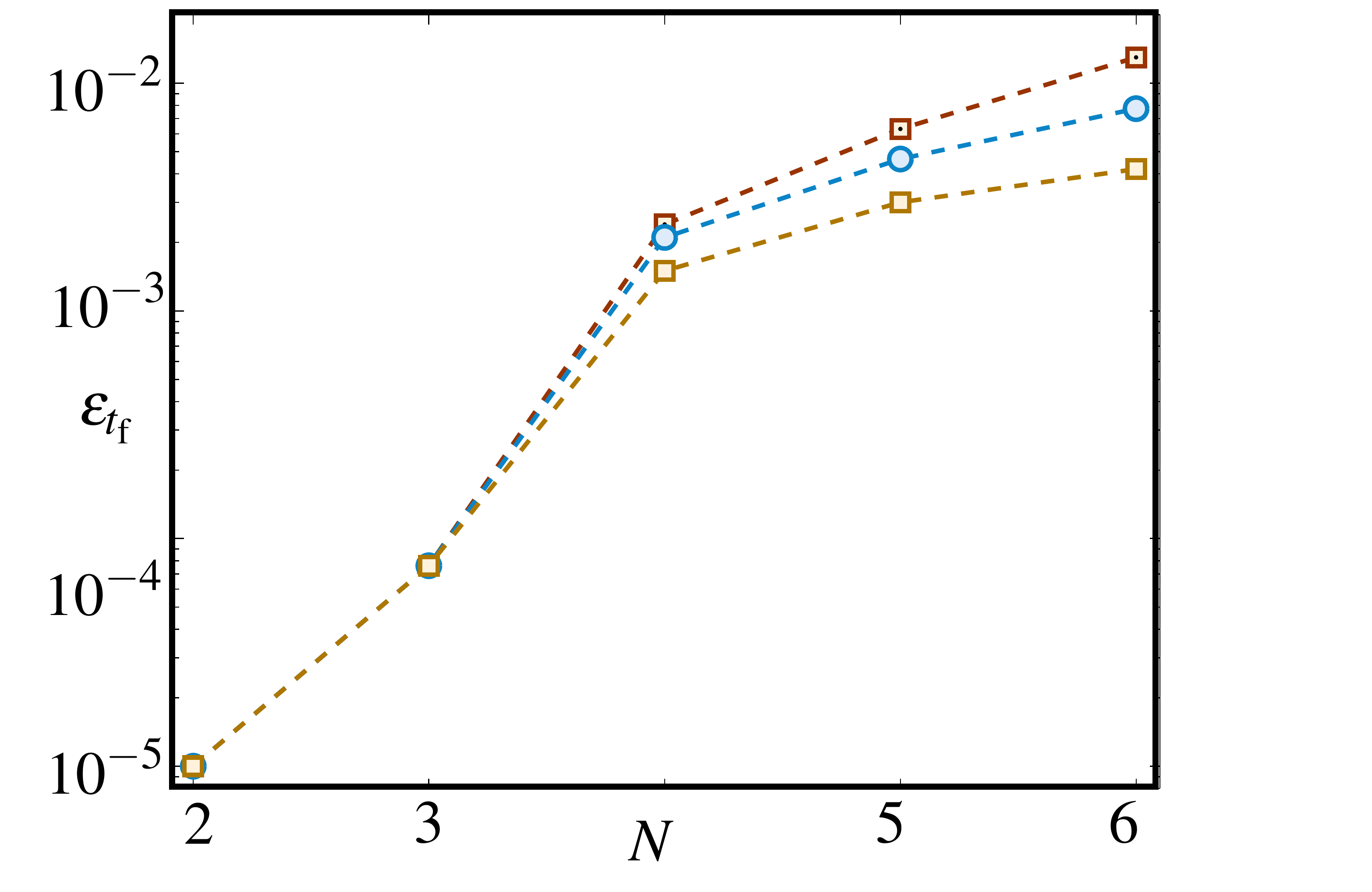}
\caption{    
\label{fig:XX} {\bf Scaling of the ground-state-cooling protocol:} Error $\epsilon_{t_{\rm f}}$ in producing the desired
ground-state $\ket{{\rm G_s}}$ after a fixed
time $t_{\rm f}=10^3/J$ and for a number of spins ranging from $N=2$ to $N=6$. For each
number of sites, we plot in blue circles the average over the different possible numbers of excitations ($n_{\rm
s}\in[1,N-1]$), and with yellow and red squares the best and worst cases, respectively. For the numerical calculation,
the Hilbert space of the oscillator was truncated to three levels.}
\end{figure}

\begin{figure*}

\centering
\includegraphics[width=1.8\columnwidth]{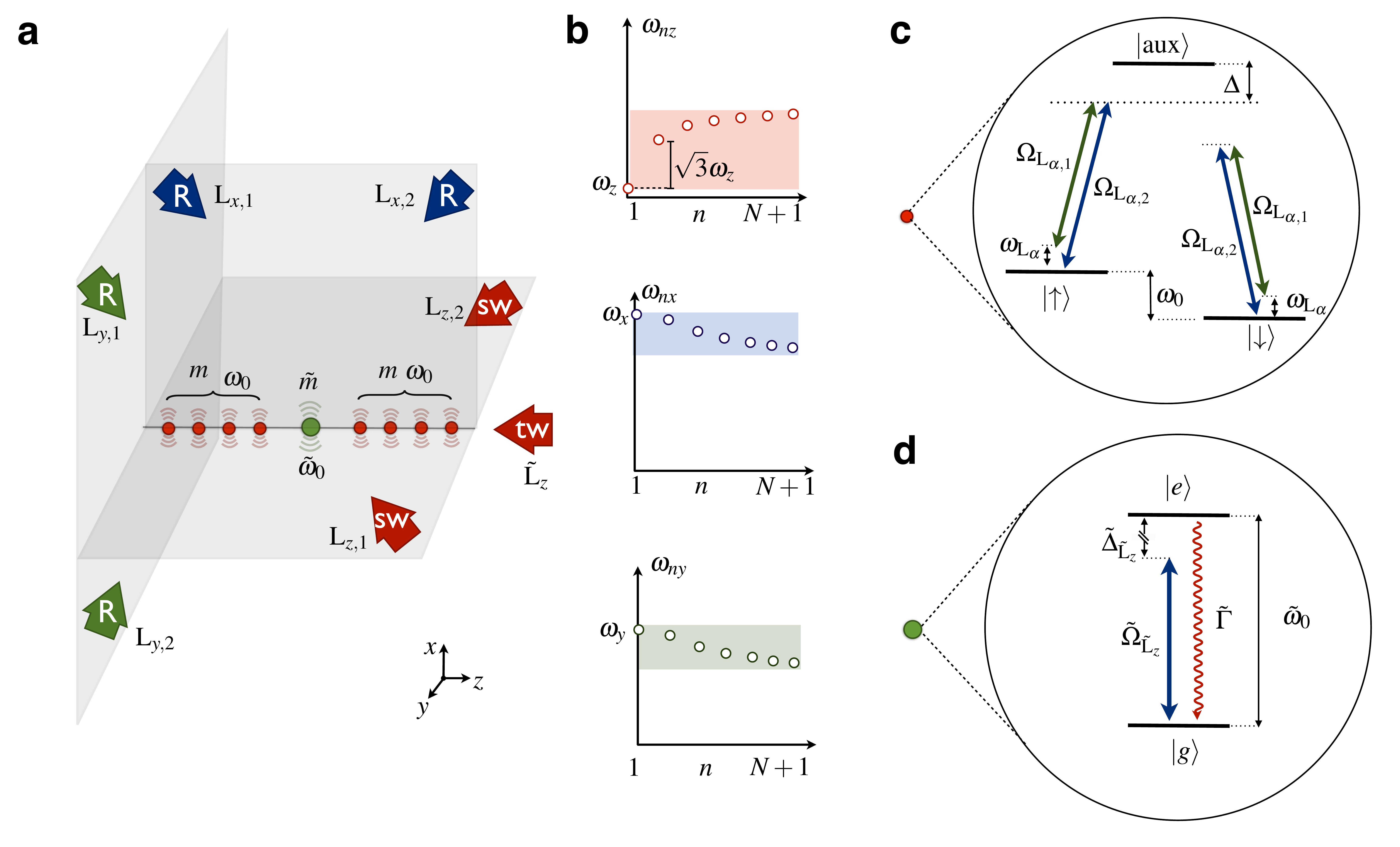}
\caption{ {\bf Trapped-ion implementation:} {\bf (a)} Mixed Coulomb crystal in a linear radio-frequency
trap. We consider two different types of ions with masses $m,\tilde{m}$, either of a different species $m\neq\tilde{m}$
or a different isotope $m\approx \tilde{m}$. The crucial property is that each type has a very different transition
frequency $\omega_0\neq\tilde{\omega}_0$. Colored arrows represent the combination of  beams in a different
configuration (i.e. R= Raman, tw=traveling
wave, sw=standing wave) leading to the desired laser-ion interaction. Let us note that the standing-wave state-dependent
forces can also be replaced by traveling waves without compromising severely the fidelities of the dissipative scheme.
{\bf (b)} Scheme of the different phonon branches for a two-isotope Coulomb crystal. The vibrational frequencies
$\omega_{n\alpha}$ span around the trap frequency of each axis $\omega_{\alpha}$. 
We note that the energy gap between the longitudinal center-of-mass mode and its neighboring mode is equal to
$(\sqrt{3}-1)\omega_z$ for the case of ions with equal masses.
{\bf (c)} Atomic $\Lambda$-scheme for the $N$ ions with hyperfine structure. A couple of laser beams with Rabi
frequencies $\Omega_{{\rm L}_{\alpha,1}},\Omega_{{\rm L}_{\alpha,2}}$ connect the two hyperfine states
$\{\ket{{\uparrow}},\ket{{\downarrow}}\}$ to an auxiliary excited state $\ket{\rm aux}$. When $\omega_{{\rm
L}_{\alpha}}\approx\omega_0$, the lasers drive a two-photon Raman transition, and a pair of Raman beams of this type
lead to the state-dependent forces in Eq.~\eqref{xy_forces}.  When $\omega_{{\rm
L}_{\alpha}}\approx\omega_{nz}\ll\omega_0$, the lasers lead to a differential ac-Stark shift, such that the contribution
from the crossed beams  leads to the state-dependent force in Eq.~\eqref{z_force}. {\bf (d)} Atomic scheme for the
$(N+1)$-th ion. A driving of the transition with Rabi frequency $\tilde{\Omega}_{{\rm L}_\alpha}$ is red-detuned from
the atomic transition $\tilde{\omega}_{{\rm L}_\alpha}=\tilde{\omega}_0+\tilde{\Delta}_{{\rm L}_{\alpha}}$, such that
$\Delta_{{\rm
L}_\alpha}<0$, leading to an effective damping of the modes.  }
\label{fig_ion_scheme}
\end{figure*}

Let us now confirm the validity of the above argument by numerical analysis for small spin chains with different numbers
of initial excitations.
In Fig. \ref{fig:XX}, we display the error in producing the ground-state~\eqref{gs_cooling} taking a fixed protocol time
of $t_{\rm f}=10^3/J$. The highest errors occur when the
number of excitations is 1 or $N-1$, since the transition frequencies are lowest at these points (this is the worst
case, with the errors 
 plotted in Fig. \ref{fig:XX} by means of red squares). With yellow squares, we represent the lowest attained
errors. Note that the average errors of the protocol (blue circles,
averaged over all the possible excitation numbers) are as low as $10^{-2}$ for chains up to $N=6$ spins, which would
thus allow us to prepare dissipatively the ground state of the spin model with fidelities of $99\%$.
 
 Let us close this section by noting
that, in the general case with an arbitrary number of spin excitations, a different choice for the site-dependence of
the 
coupling coefficients $g_i$ may allow for jump operators that go down/up the spin ladder in larger steps, and therefore
favor a faster convergence to the asymptotic state (this appears to be specially well-suited for chains around the
half-filling condition). However, for up to $N=6$ spins, we have not found any sizable advantage, and the highest
fidelities were always achieved for transitions between neighboring spin waves $n\to n\pm1$.

\section{Realization with crystals of trapped ions}
\label{sec:ions}

\subsection{ Trapped-ion crystal as a spin-boson chain} 
\label{ion_crystal}

Let us describe a specific platform where the DSBC model~\eqref{dsbc} can be realized. We consider a system of $N+N'$
ions
confined in a linear radio-frequency trap~\cite{wineland_review}, and arranged forming a one-dimensional chain (see
Fig.~\ref{fig_ion_scheme}{\bf (a)}). The vibrational degrees of freedom around the equilibrium positions can be
expressed in terms of the so-called normal modes~\cite{james}, namely
\begin{equation}
H_{\rm p}=\sum_{\alpha,n}\omega_{n\alpha}b_{n\alpha}^{\dagger}b_{n\alpha}^{\phantom{\dagger}},
\end{equation} 
where $\alpha\in\{x,y,z\}$ refers to the trap axis, $n\in\{1,\dots,N+N'\}$ labels the different modes with frequencies
$\omega_{n\alpha}$, and $b_{n\alpha}^{\phantom{\dagger}},b_{n\alpha}^{{\dagger}},$ are bosonic operators that
annihilate/create phonons in a particular mode. In a linear chain of ions with equal mass, the phonon branches have the
structure shown in
Fig.~\ref{fig_ion_scheme}{\bf (b)}, such that the different modes spread around the trap frequencies lying in the range 
$\omega_x/2\pi,\approx$1-10MHz, $\omega_y/2\pi,\approx$1-10MHz, and $\omega_z/2\pi\approx$0.1-1MHz. A property of the
longitudinal branch that will be used in this work is the existence of an energy gap between the lowest-energy mode
(i.e. the center-of-mass mode) and the following one. 

As shown in Fig.~\ref{fig_ion_scheme}{\bf (a)}, $N$ ions of the chain correspond to a particular atomic species with
hyperfine structure, whereas the remaining $N'$ ions, that will be used for cooling, do not necessarily have 
hyperfine splitting (we note that though we focus on hyperfine qubits, our proposal could be generalized to optical
or Zeeman qubits). For simplicity, in the following we describe the case $N'=1$. For the first $N$
ions, we select two hyperfine levels of the ground-state manifold, which are referred to as spin states
$\{\ket{{\uparrow_i}},\ket{{\downarrow_i}}\}$, and have a transition frequency $\omega_0$ in the microwave range with a
negligible linewidth $\Gamma\approx 0$ (Fig.~\ref{fig_ion_scheme}{\bf (c)}). For the remaining $(N+1)$-th ion, we select
two internal states $\{\ket{{g}},\ket{{e}}\}$ from a certain transition with frequency $\tilde{\omega}_0$, and linewidth
$\tilde{\Gamma}/2\pi\approx$10-100$\,$kHz (Fig.~\ref{fig_ion_scheme}{\bf (d)}). This is the so-called resolved sideband
limit $\tilde{\Gamma}\ll\omega_z$, which will allow for an efficient laser cooling close to the vibrational ground state
for the longitudinal center-of-mass mode \cite{symp_sideband_cooling}. Let us note that these small linewidths 
can be obtained by a variety of methods depending on the particular ion species (e.g. for weakly-allowed dipole
transitions, for quadrupole-allowed transitions with an additional laser that admixes the excited state with that of a
dipole-allowed transition, or by a Raman configuration)~\cite{leibfried_review}.

The complete degrees of freedom are thus described by the following master equation
\begin{equation}
\label{meq}
\dot{\rho}={\mathcal{L}}(\rho)=-\ii[H_{\tau}+\tilde{H}_{\tau}+H_{\rm p},\rho]+\tilde{\mathcal{D}}(\rho),
\end{equation}
where we have introduced the Hamiltonians
\begin{equation}
H_{\tau}=\sum_{i=1}^{N}\half\omega_0
\tau_i^z,\hspace{2ex}\tilde{H}_{\tau}=\half{\tilde{\omega}_0}\tilde{\tau}_{N+1}^z \,,
\end{equation}
such that $\tau_i^z=\ketbra{{\uparrow_i}}-\ketbra{{\downarrow_i}}$, and
$\tilde{\tau}_{N+1}^z=\ketbra{{e}}-\ketbra{{g}}$ (we use $\tau$ instead of $\sigma$ to
emphasize the pseudo-spin character of these internal states). Additionally, we have to take into account the
dissipation from the
dipole-allowed transition considering recoil effects~\cite{recoil_master_equation}. For trapped ions, this can be
described by the following super-operator
$\tilde{\mathcal{D}}(\rho)=\tilde{\mathcal{D}}^0(\rho)+\tilde{\mathcal{D}}^1(\rho)$, where
\begin{equation}
\label{do}
\tilde{\mathcal{D}}^0(\bullet)=\frac{\tilde{\Gamma}}{2}\big(\tilde{\tau}_{N+1}^-\bullet
\tilde{\tau}_{N+1}^+-\tilde{\tau}_{N+1}^+\tilde{\tau}_{N+1}^-\bullet+\text{H.c.}\big),
\end{equation}
where $\tilde{\tau}_{N+1}^+=\ket{e}\bra{g}=(\tilde{\tau}_{N+1}^-)^{\dagger}$. This is the usual dissipator of a
two-level atom in an electromagnetic environment. The spontaneous emission also has effects on the
vibrational degrees of freedom due to the photon recoil. To leading order in the Lamb-Dicke parameter
$\eta_{n\alpha}=(\tilde{\omega}_0/c)/\sqrt{2\tilde{m}\omega_{n\alpha}}$ for each normal mode, the jump operators
create and annihilate phonons as given by \cite{laser_cooling_trapped_ions}:
\begin{equation}
\label{d1}
\begin{split}
\tilde{\mathcal{D}}^1(\bullet)\!=\!\!\!\sum_{\alpha,n}\!\half\tilde{\Gamma}_{n\alpha}\tilde{\tau}_{N+1}^-\!&\big[
\!(b_{n{\alpha}}^{\dagger}+b_{n{\alpha}}^{\phantom{\dagger}})\bullet(b_{n{\alpha}}^{\dagger}+b_{n{\alpha}}^{\phantom{
\dagger}})\\
&-(b_{n{\alpha}}^{\dagger}+b_{n{\alpha}}^{\phantom{\dagger}})^2\bullet\big]\tilde{\tau}_{N+1}^++\text{H.c.},
\end{split}
\end{equation}
where we have introduced
$\tilde{\Gamma}_{n\alpha}\propto\tilde{\Gamma}_{\alpha}\tilde{\eta}_{n{\alpha}}^2(\mathcal{M}_{N+1,n}^{
\alpha }
)^2$, with $\mathcal{M}_{N+1,n}^{\alpha}$ the coefficients describing the displacement of
the $N+1$ ion in mode $n$ along direction $\alpha$. 

Once the vibrational and atomic degrees of freedom have been introduced, let us  summarize the role that each of them
will play in the realization of the damped spin-boson chain~\eqref{dsbc}: {\it (i)} The hyperfine levels of the $N$ ions
will simulate the spins in the DSBC model. {\it (ii)} The $2(N+1)$ vibrational modes along the $x$ and $y$ axes will be
used to mediate the spin-spin interactions leading to the isotropic XY model in Eq.~\eqref{xy_model}. {\it (iii)} The
longitudinal center-of-mass mode will play the role of the harmonic oscillator in the DSBC model. {\it (iv)} The
atomic levels of the $(N+1)$-th ion will be used to laser-cool the longitudinal center-of-mass mode.  We note that the
position of
this auxiliary ion is irrelevant for the purpose of cooling, but it modifies the spin couplings in the chain. Moreover,
it is also possible to use more than one auxiliary ion as an additional knob to control the damping rate. 
In the following sections, we will detail the different elements required for the ion-trap
implementation of the DSBC. We emphasize that all
of them can be realized with state-of-the-art technology.

\subsection{Engineering the isotropic XY spin model}
\label{ion_xy_model}

We start by addressing how to implement the isotropic XY model~\eqref{xy_model} in the ion chain by means of a variation
of the M{\o}lmer-S{\o}rensen gate \cite{Molmer-Sorensen}. The idea is to use the so-called spin-dependent forces, a tool
that has already been implemented in several laboratories for different
purposes~\cite{state_dependent_forces_cats,state_dependent_forces_gates,state_dependent_forces_simulations,
state_dependent_forces_walks,state_dependent_forces_mw}. By combining a pair of Raman laser beams (see
Fig.~\ref{fig_ion_scheme}{\bf (c)}), each of which is tuned to the red/blue first vibrational sideband of the hyperfine
transition~\cite{state_dep_review}, it is possible to obtain the following state-dependent force
\begin{equation}
\label{xy_forces}
\begin{split}
\hat{H}_{\rm
L_{\alpha}}\!\!&=\!\sum_{in}\!\mathcal{F}^{\alpha}_{in}\tau_i^{\phi_{\alpha}}b_{n\alpha}\ee^{\ii\varphi_\alpha}\ee^{
-\ii\delta_{n\alpha}t}+\text{H.c.},\\
\mathcal{F}_{in}^{\alpha}&=\ii\half\eta_{n\alpha}\Omega_{\rm L_{\alpha}}\mathcal{M}_{in}^{\alpha},
\end{split}
\end{equation}
where the ``hat'' indicates the interaction picture with respect to $H_0=H_{\sigma}+\tilde{H}_{\sigma}+H_{\rm p}$. In
the
above expression, we have assumed that the effective wavevector of the Raman beams ${\bf k}_{\rm L_\alpha}={\bf k}_{{\rm
L}_{\alpha,1}}-{\bf k}_{{\rm L}_{\alpha,2}}$ points along the $\alpha=\{x,y\}$ axis of the trap (see
Fig.~\ref{fig_ion_scheme}{\bf (a)}). Additionally, we assume that the strength of both Raman beams is equal, such that
they share a common Rabi frequency $|\Omega_{\rm L_{\alpha}}|$, and have opposite detunings, such that we can define
a common $\delta_{n\alpha}=\omega_{{\rm L}_\alpha}-(\omega_0-\omega_{n\alpha})$. We have introduced the laser Lamb-Dicke
parameter $\eta_{n\alpha}={\bf e}_{\alpha}\cdot{\bf k}_{\rm L_\alpha}/\sqrt{2m\omega_{n\alpha}}\ll1$.  The constraints
on these parameters are $|\Omega_{\rm L_\alpha}|\ll\omega_{\alpha}$ to neglect other terms in addition to such a
state-dependent force. Finally, we have defined the sum and difference of the Raman beam phases 
\begin{equation}
\phi_\alpha=\phi_{{\rm L}_{\alpha,1}}+\phi_{{\rm L}_{\alpha,2}},\hspace{2ex} \varphi_\alpha=\phi_{{\rm
L}_{\alpha,1}}-\phi_{{\rm L}_{\alpha,2}} \,,
\end{equation}
 and the Pauli spin operator
$\tau_i^{\phi_\alpha}=\ee^{-\ii\phi_\alpha}\ket{{\uparrow_i}}\bra{{\downarrow_i}}+\ee^{+\ii\phi_\alpha}\ket{{
\downarrow_i}}\bra{{\uparrow_i}}$.

In order to obtain the desired isotropic XY model, we apply two orthogonal state-dependent forces with the following
phases and directions
\begin{equation}
\alpha=x,\hspace{1ex}\phi_x=0,\hspace{2ex}\alpha=y,\hspace{1ex}\phi_y=\textstyle{\frac{\pi}{2}}.
\end{equation}
Altogether, the Hamiltonian of the ion chain in the interaction picture becomes
\begin{equation}
\hat{H}_{\rm
L}\!\!=\!\sum_{in}\!\mathcal{F}^{x}_{in}\tau_i^{x}b_{nx}\ee^{\ii\varphi_x}\ee^{-\ii\delta_{nx}t}+\sum_{in}\!\mathcal{F}^
{y}_{in}\tau_i^{y}b_{ny}\ee^{\ii\varphi_y}\ee^{-\ii\delta_{ny}t}+\text{H.c.}\,.
\end{equation}

Performing a Magnus expansion~\cite{magnus}, the evolution turns out to be given by the unitary operator
$\hat{U}(t)=\exp\{ \Omega_{\rm eff}(t)\}$, where $\Omega_{\rm eff}(t)$ displays the following  form to second order
\begin{equation}
\Omega_{\rm eff}(t) = -\ii\!\! \int_0^t\!\!\!\!  {\rm d}t' \tilde H_{\rm L}(t') \\ - \frac{1}{2 }\!  \int_0^t \!\!\!\! 
{\rm d}t'\!\!\!\!  \int_0^{t'} \!\! \!\! {\rm d}t'' [\tilde H_{\rm L}(t'), \tilde H_{\rm L}(t'')].
\end{equation}
We want to derive an effective Hamiltonian from this expression $\Omega_{\rm eff}(t)\approx-\ii t H_{\rm s}^{\rm ti}$
(where the superscript ``ti'' stands for the trapped-ions realization).
The first-order contribution leads to a couple of orthogonal state-dependent displacements, which can only be neglected
in the limit $|\mathcal{F}_{in}^{\alpha}|\ll\delta_{n\alpha}$~\cite{porras_spins}. In addition, from the
non-commutativity of the $\sigma^x$ and $\sigma^y$ forces, we get a residual spin-phonon coupling in the second-order
term of the Magnus expansion. In order to neglect it, we have to impose a further constraint, namely
$|\mathcal{F}_{in}^x(\mathcal{F}_{im}^y)^*|\ll|\delta_{nx}-\delta_{my}|$, which can be fulfilled if the trap frequencies
along the $x,y$ axes are sufficiently different $\omega_x\neq \omega_y$. Under these constraints, an interacting quantum
spin chain is obtained
\begin{equation}
\label{spin_model}
H_{\rm s}^{\rm
ti}=\sum_{i,j}J_{ij}^x\tau_i^x\tau_j^x+J_{ij}^y\tau_i^y\tau_j^y,\hspace{2ex}J_{ij}^{\alpha}=-\sum_n\frac{1}{\delta_{
n\alpha}}\mathcal{F}_{in}^\alpha(\mathcal{F}_{jn}^\alpha)^*.
\end{equation}
Therefore, by adjusting the strengths of the Raman beams, and their detunings, it is possible to find a regime where
$J_{ij}^x=J_{ij}^y$, and the above spin Hamiltonian corresponds to the desired isotropic XY model since $\tau_i^x
\tau_j^x+ \tau_i ^y \tau_j^y = 2(\tau_i ^+ \tau_j^- + \tau_i ^-\tau_j^+)$. Let us remark that the resulting XY model has
the
peculiarity of displaying long-range couplings. In fact, when $\omega_x,\omega_y\gg\omega_z$, and the Raman lasers are
far-detuned from the whole vibrational branch, it can be shown that these couplings decay with a dipolar law.
The trapped-ion Hamiltonian~\eqref{spin_model} becomes a realization of the XY
interaction~\eqref{xy_model} in the DSBC model~\eqref{dsbc} after the identifications
$\tau_i^{\pm}\leftrightarrow\sigma_i^{\pm}$, and $J_{ij}^{x}=J_{ij}^{y}\leftrightarrow J$. For typical nearest-neighbor
distances of $z_0\approx$1-10$\mu$m, the spin-spin couplings attain
strengths in the $J_{ii+1}^{\alpha}/2\pi\approx$1-10 kHz. Let us note that
modifications of the scheme of state-dependent forces have been proposed to yield other quantum spin
models~\cite{porras_spins,effective_spin_models_theory}, some of which have also been realized
experimentally~\cite{state_dependent_forces_simulations}.

\subsection{ Controlling the spin-boson coupling} \label{subsec:dipole force} 
\label{ion_spin_boson}

Let us now address how to implement the spin-boson coupling~\eqref{spin_boson} in the ion chain. The idea is again to
use a $\Lambda$-beam configuration (see Fig.~\ref{fig_ion_scheme}{\bf (c)}), but in a different regime. Rather
than tuning the two-photon frequencies $\omega_{\rm L_{\alpha}}$ to the vibrational sidebands of the hyperfine
transition, we impose  that $\omega_{\rm L_{\alpha}}\approx\omega_{nz}\ll \omega_0$. Moreover, we consider that the
laser beams form a standing wave  along the trap $z$-axis (see Fig.~\ref{fig_ion_scheme}{\bf (a)}). Under these
constraints, the laser-ion interaction leads to a crossed-beam differential ac-Stark shift, which can be interpreted as
another state-dependent force in the $\tau^z$ basis
\begin{equation}
\label{z_force}
\begin{split}
\hat{H}_{\rm L_{z}}\!\!&=\!\sum_{in}\!\mathcal{F}^{z}_{in}\tau_i^{z}b_{nz}\ee^{-\ii\delta_{nz}t}+\text{H.c.},\\
\mathcal{F}_{in}^{z}&=\half\eta_{nz}\Omega_{{\rm L}_{z}}\mathcal{M}_{in}^{z}\sin(\varphi_{z}-{{\bf k}_{{\rm
L}_z}\cdot{\bf r}_i^0}),
\end{split}
\end{equation}
where the ``hat'' refers to  the interaction picture with respect to $H_0=H_{\tau}+\tilde{H}_{\tau}+H_{\rm p}$.
Here,
the parameters are defined in analogy to those in Eq.~\eqref{xy_forces}, with three important differences: {\it (i)}
$\Omega_{{\rm L}_{z}}$ is not the two-photon Rabi frequency of the hyperfine transition, but rather a differential
ac-Stark shift coming from processes where a photon is exchanged between the pair of laser beams in the $\Lambda$
configuration. {\it (ii)} The detunings are changed to $\delta_{nz}=\omega_{{\rm L}_z}-\omega_{nz}$.
{\it (iii)} Since the ion chain lies along the $z$ axis, there is an additional site-dependent phase when shining the
lasers such that  ${\bf k}_{{\rm L}_z}\cdot{\bf r}_i^0\neq 0$. When adjusting the difference of the laser phases to be
$\varphi=\pi/2$, then we obtain $\mathcal{F}_{in}^{z}=\half\eta_{nz}\Omega_{{\rm L}_{z}}\mathcal{M}_{in}^{z}\cos({{\bf
k}_{{\rm L}_z}\cdot{\bf r}_i^0})$. Let us highlight that the standing-wave nature of this spin-dependent force is not
essential for the dissipative protocol. However, it makes the connection with the DSBC studied in
Sec.~\ref{dissipative_protocol} more transparent. We will show in Sec.~\ref{numerical} that essentially the same
fidelities can be achieved for a traveling-wave configuration, which has also been experimentally
demonstrated~\cite{state_dep_review}.

We now exploit the particular properties of the longitudinal phonon branch (see Fig.~\ref{fig_ion_scheme}{\bf (b)}).
More precisely, we use the presence of a gap between the axial center-of-mass mode and the rest of the modes along the
same direction. For
the case of equal ions, $|\omega_{nz}-\omega_{1z}|\geq (\sqrt{3}-1)\omega_z$, while for different ion species or
isotopes, the exact value of the gap will depend on the mass ratio and the position of the cooling ions. We 
assume that $\omega_{{\rm L}_z}$ is red-detuned with respect to the  axial center-of-mass mode
$\omega_{{\rm L}_z}=\omega_{1z}-\delta_{1z}$, such that $\delta_{1z}\ll\omega_{1z}$. Since the next vibrational modes
are separated by a large energy gap, the laser coupling  to the
remaining phonon branch is highly off-resonant and can be thus neglected. Accordingly,
this state-dependent force gives
\beq
 \hat{H}_{{\rm L}_z}=\sum_i\mathcal{F}_{i1}^z\tau_i^zb_{1z}\ee^{-\ii\delta_{1z}t}+ {\rm H.c.}.
\eeq
We can finally move to a different picture where the above Hamiltonian becomes time-independent $H_{{\rm L}_z}=H_{\rm
b}^{\rm ti}+H_{\rm sb}^{\rm ti}$, where
\beq
\label{spin_boson_ion}
\begin{split}
 {H}_{{\rm b}}^{\rm ti}&=\sum_i\delta_{1z}b_{1z}^{\dagger}b_{1z},\\
 H_{\rm sb}^{\rm ti}&=\sum_i\mathcal{F}_{i1}^z\tau_i^z(b_{1z}+b_{1z}^{\dagger}).
 \end{split}
\eeq
Therefore, it is clear that the center-of-mass mode plays the role of the harmonic oscillator
$\{b_{1z},b_{1z}^{\dagger}\}\leftrightarrow \{a,a^{\dagger}\}$ in the DSBC, with the laser detuning corresponding to the
oscillator detuning $\delta_{1z}\leftrightarrow\Delta_a$ in the DSBC model~\eqref{oscillator}. In addition, the strength
of the state-dependent force determines the spin-boson coupling $\mathcal{F}_{in}^{z}\leftrightarrow g_i$ in the
DSBC model~\eqref{spin_boson}.

Let us now comment on realistic values for the trapped-ion parameters. Since the laser detuning should only fulfill
$\delta_{1z}/2\pi\ll\omega_{1z}/2\pi=\omega_z/2\pi\approx$1-10$\,$MHz, it will be easy to reach the required condition
of $\delta_{1z}/2\pi\sim J_{ii+1}^{\alpha}/2\pi\approx$1-10$\,$kHz. On the other hand, we know from the previous section
that $g\ll J$ is necessary to have an accurate pumping to the desired entangled state. Nevertheless, $g$ should not be
too small that
the total preparation time becomes prohibitively long. A
suitable choice could be $g\approx$ 0.1-0.5 kHz. Since $\mathcal{F}_{i1}^{z}=\eta_{1z}\Omega_{{\rm L}_{z}}\cos({{\bf
k}_{{\rm L}_z}\cdot{\bf r}_i^0})/\sqrt{N+1}$, it will suffice to set the Rabi frequency in the $\Omega_{{\rm
L}_{z}}\approx$1-5$\,\sqrt{N}\,$kHz. Finally, we should adjust the laser wavevector such that ${\bf k}_{{\rm
L}_z}\cdot{\bf r}_i^0\approx \pi i/(N+1)$.

\subsection{ Effective damping of the bosonic mode}
\label{ion_damping}

Once the implementation of the  XY model~\eqref{xy_model}  and the spin-boson coupling~\eqref{spin_boson} has been
described, let us turn into the last required ingredient: an effective damping of the bosonic mode~\eqref{damping}. As
discussed previously, the idea is to exploit the atomic levels of the $(N + 1)$-th ion to laser-cool the longitudinal
center-of-mass mode. Since the resonance frequencies are very different, $\omega_0\neq\tilde{\omega}_0$, the cooling
lasers do not affect the spin dynamics of the other $N$ ions. However, since the vibrational modes are collective, it is
possible to sympathetically cool the vibrations of the crystal by only acting on the $(N + 1)$-th ion. This sympathetic
laser
cooling~\cite{symp_cooling_theory} has been already realized experimentally in small crystals
for quantum computation~\cite{symp_sideband_cooling, sympathetic_cooling}.

Our starting point is the master equation~\eqref{meq} with the dissipation superoperators~\eqref{do}-\eqref{d1}. To
control the effective damping, we introduce a laser beam red detuned from the atomic transition
$\tilde{\omega}_{\tilde{\rm L}_z}\approx\tilde{\omega}_0+\tilde{\Delta}_{\tilde{\rm L}_z}$, such that
$\tilde{\Delta}_{\tilde{\rm L}_z}<0$ (see Fig.~\ref{fig_ion_scheme}{\bf (d)}). Let us note that the particular
laser-beam configuration will depend on the particular ion species, and the scheme to attain the resolved-sideband
limit~\cite{leibfried_review}.
We consider that the laser is in a traveling-wave configuration, such that its wavevector is aligned parallel to the
trap axis $\tilde{\bf k}_{{\rm L}_z}\parallel{\bf e}_z$ (see Fig.~\ref{fig_ion_scheme}{\bf (a)}). After expanding in
series of the Lamb-Dicke parameter $\tilde{\eta}_{nz}={\bf e}_{z}\cdot\tilde{{\bf k}}_{\tilde{\rm
L}_z}/\sqrt{2\tilde{m}\omega_{nz}}\ll1$, the laser-ion interaction leads to two different terms, the so-called carrier
term
\begin{equation}
\tilde{H}_{\tilde{\rm L}_{z}}^0=\half\tilde{\Omega}_{\tilde{\rm L}_z}\ee^{\ii\tilde{\bf k}_{\tilde{\rm L}_z}\cdot {\bf
r}_{N+1}^0}\tilde{\tau}_{N+1}^+\ee^{-\ii\tilde{\omega}_{\tilde{\rm L}_z}t}+{\rm H.c.} \,,
\end{equation} 
and the red and blue sideband terms
\begin{equation}
\tilde{H}_{\tilde{\rm
L}_{z}}^1=\sum_{n}\tilde{\mathcal{F}}_{N+1,n}^z(b_{nz}+b_{nz}^{\dagger})\tilde{\tau}_{N+1}^+\ee^{-\ii\tilde{\omega}_{{
\rm L}_z}t}+{\rm H.c.},
\end{equation} 
where we introduced $ \tilde{\mathcal{F}}_{N+1,n}^{z}=\ii\half\tilde{\eta}_{nz}\tilde{\Omega}_{\tilde{\rm
L}_{z}}\mathcal{M}_{N+1,n}^{z}\ee^{\ii\tilde{\bf k}_{\tilde{\rm L}_z}\cdot {\bf r}_{N+1}^0}$, and the corresponding Rabi
frequency $\tilde{\Omega}_{\tilde{\rm L}_{z}}$. By controlling appropriately these sideband terms, and their interplay
with the atomic spontaneous emission, it is possible to tailor the damping of the longitudinal modes. 

Let us rearrange the full Liouvillian~\eqref{meq}  as a sum of two terms
${\mathcal{L}}={\mathcal{L}}_0+{\mathcal{L}}_1$, where
\begin{equation}
\begin{split}
\mathcal{L}_0(\bullet)&=-\ii[H_s+\tilde{H}_s+H_{\rm b}+\tilde{H}_{\tilde{\rm
L}_z}^0,\bullet]+\tilde{\mathcal{D}}^0(\bullet),\\
\mathcal{L}_1(\bullet)&=-\ii[\tilde{H}_{\tilde{\rm L}_z}^1,\bullet]+\tilde{\mathcal{D}}^1(\bullet),
\end{split}
\end{equation}
which is justified for small Lamb-Dicke parameters. To obtain the effective damping of the 
 longitudinal center-of-mass mode, we use a similar formalism as in \cite{ad_elim}. In this case,
the fastest timescale in the problem
is given by the decay rate $\tilde{\Gamma}$ of the atomic states of the $(N+1)$-th ion. Therefore, we must  ``integrate
out'' these atomic  degrees
of freedom, which can be accomplished by projecting the
density matrix of the full system into  $\tilde{\mathcal{P}}\rho(t)=\rho_{N+1}^{\rm
ss}\otimes\rho_{N}(t)$, where $\rho_{N+1}^{\rm ss}$ fulfills $\tilde{\mathcal{D}}^0(\rho_{N+1}^{\rm ss})=0$, and
$\rho_{N}(t)$ is the reduced density matrix for the $N$ spins and the chain vibrational modes. For this we use the
expression
\begin{equation}
\frac{{\rm d}\hat{\rho}_{N}}{{\rm d}t}=\int_0^{\infty}{\rm d}\tau{\rm
Tr}_{N+1}\left\{\tilde{\mathcal{P}}\hat{\mathcal{L}}_1(t)\ee^{\hat{\mathcal{L}}_0\tau}\hat{\mathcal{L}}_1(t-\tau)\tilde{
\mathcal{P}}\hat{\rho}(t)\right\},
\end{equation}
where the ``hats'' refer to the  interaction picture with respect to $H_0=H_{\tau}+\tilde{H}_{\tau}+H_{\rm p}$.
After
some algebra, one gets the effective damping of the longitudinal modes
\begin{equation}
\dot{\rho}_N=-\ii[H_{\rm s}+H_{\rm b},\rho_N]+{\mathcal{D}}_{\rm b}(\rho_{N}),
\end{equation}
where we have introduced the bosonic dissipator
\begin{equation}
\label{eff_cooling}
{\mathcal{D}_{\rm b}}(\bullet)\!=\!\!\sum_n\!\kappa_n^-(b_{nz}^{\phantom{\dagger}}\bullet
b_{nz}^{\dagger}\!-b_{nz}^{\dagger}b_{nz}^{\phantom{\dagger}}\bullet)+\kappa_n^+(b_{nz}^{{\dagger}}\bullet
b_{nz}^{\phantom{\dagger}}\!-b_{nz}^{\phantom{\dagger}}b_{nz}^{{\dagger}}\bullet)+{\rm H.c.} \,,
\end{equation}
together with the effective rates for laser cooling and heating 
\begin{equation}
\kappa_n^{\mp}=D_n+S_n(\pm\omega_{nz}).
\end{equation}
Here, the so-called diffusion coefficient accounting for the recoil heating is
$D_n=\tilde{\Gamma}_{nz}\langle\tilde{\tau}^+_{N+1}\tilde{\tau}^-_{N+1}\rangle_{\rm ss}$, and the spectral functions
of the sideband terms are
\begin{equation}
S_n(\omega)=\int_0^{\infty}{\rm d}\tau\langle(\hat{F}_n(\tau) \hat{F}_n(0)\rangle_{\rm
ss}\ee^{\ii\omega\tau},\hspace{1ex}F_n=\tilde{\mathcal{F}}^z_{N+1,n}\tau_i^++{\rm H.c.} \,,
\end{equation}
which can be obtained by means of the quantum regression theorem. The formalism and cooling rates coincide with those of
a single trapped ion~\cite{laser_cooling_trapped_ions}, with the difference that in the diffusion coefficient $D_n$ and
forces $F_n$ one must consider the normal vibrational modes at the position of the cooled ion.

The idea is to work in the resolved sideband regime $\tilde{\Gamma}\ll\omega_z$, and  set the laser-cooling  parameters
to optimize the cooling of the  longitudinal center-of-mass mode $\kappa_1^-\gg\kappa_1^+$. In this regime, the cooling
dominates in~\eqref{eff_cooling}, and we are left with the desired damping of the bosonic mode 
\begin{equation}
\label{cm_cooling}
\mathcal{D}_{\rm b}^{\rm ti}(\bullet)\!=\!\!\kappa_1^-(b_{1z}^{\phantom{\dagger}}\bullet
b_{1z}^{\dagger}\!-b_{1z}^{\dagger}b_{1z}^{\phantom{\dagger}}\bullet)+\kappa_1^+(b_{1z}^{{\dagger}}\bullet
b_{1z}^{\phantom{\dagger}}\!-b_{1z}^{\phantom{\dagger}}b_{1z}^{{\dagger}}\bullet)+{\rm H.c.}.
\end{equation}
Once more, after the identifications $\{b_{1z},b_{1z}^{\dagger}\}\leftrightarrow \{a,a^{\dagger}\}$, and
$\kappa_1^-\leftrightarrow\kappa$, we obtain an analogous  DSBC damping~\eqref{damping}.

Let us finally comment on the required cooling rates.  Ideally, we should achieve the regime $\kappa\approx
g\approx\,$0.1-0.5 kHz. Since such cooling is not particularly fast, it is possible to find the right detuning
$\tilde{\Delta}_{\tilde{\rm L}_z}<0$ such that $\kappa_1^-\approx\,$0.1-0.5 kHz. Besides, since we are in the resolved
sideband limit, the heating can be made much smaller, $\kappa_1^+\ll\kappa_1^-$. The stationary state of a
harmonic oscillator under the action of $\mathcal{D}_{\rm b}^{\rm ti}$ in Eq.~\eqref{cm_cooling}  is a thermal state
with mean number of quanta
$\bar{n}_{1z}=\kappa^+_1/(\kappa_1^--\kappa_1^+)$. In the limit $\kappa_1^+/\kappa_1^-=\zeta\ll1$,
such that the cooling is the dominant effect and the heating only presents a small correction, the
equilibrium state of the center-of-mass mode is close to the vacuum state.

\subsection{Numerical analysis of the trapped-ion dissipative protocol}
\label{numerical}

In this last section, we will explore numerically how the dissipative protocol described in the  part~\ref
{dissipative_protocol} of this manuscript can be implemented with the trapped-ion DSCB model in
Eqs.~\eqref{spin_model},~\eqref{spin_boson_ion} and~\eqref{cm_cooling},  namely
\begin{equation}
\label{dsbc_ion}
\frac{{\rm d}\rho}{{\rm d}t}=\mathcal{L}_{\rm DSBC}^{\rm ti}(\rho)=-\ii[H_{\rm b}^{\rm ti}+H_{\rm s}^{\rm ti}+H_{\rm
sb}^{\rm ti},\rho]+\mathcal{D}_{\rm b}^{\rm ti}(\rho).
\end{equation}
Therefore,  the results discussed below correspond to the actual spin-spin couplings $J_{ij}^{\alpha}$ in an ion trap,
which are not nearest-neighbors couplings, but display a dipolar decay with the cube of the distance. Furthermore,
we have taken into account that in  harmonic ion traps, the interparticle distance is not constant over the chain. 
More importantly, we have considered the effects of the always-present heating term in the trapped-ion effective
damping~\eqref {cm_cooling}. 

\begin{figure}
\centering
 \includegraphics[width=0.9\columnwidth]{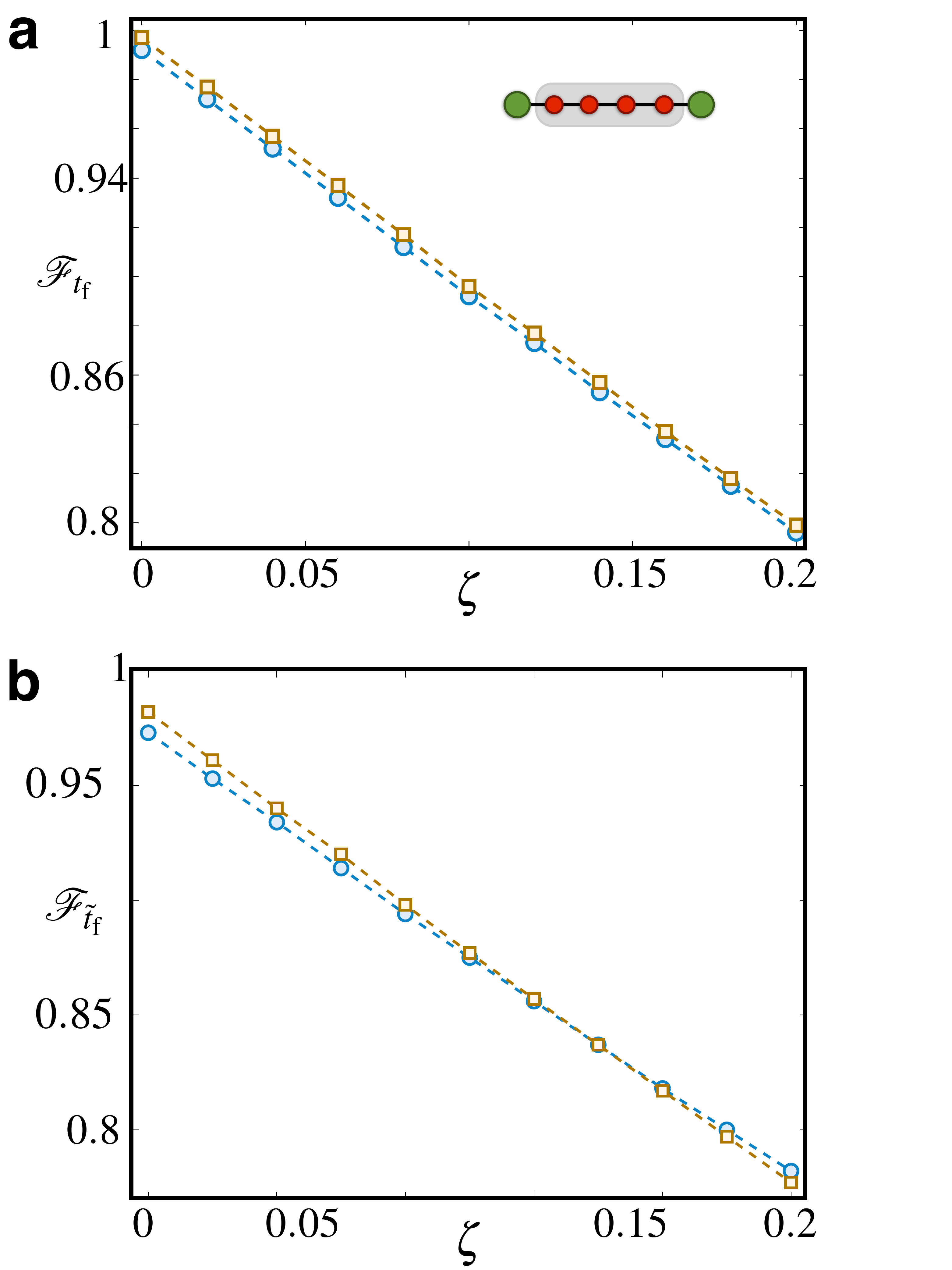}
\caption{    
\label{fig:6 ion chain} {\bf Heating in the trapped-ion dissipative protocol:} {\bf (a)} Fidelity with the desired
target state  for a
fixed
time $t_{\rm f}=10^3/J$ and as a function of the ratio $\zeta=\kappa_1^+/\kappa_1^-$ between the sympathetic heating and
cooling rates in a chain of six ions
in which the central four represent sites of the isotropic XY spin chain. The blue dots correspond to the case of
$n_{\rm s}=\{1,3\}$ initial spin excitations, and the orange squares to the case of $n_{\rm s}=2$ excitations. The
oscillator was
truncated to five levels. {\bf (b)} Same as in {\bf (a)}, but considering a shorter time  $\tilde{t}_{\rm
f}=10^2/J$, and a traveling-wave configuration of the spin-boson coupling in Eq.~\eqref{spin_boson_ion}. The achieved
fidelities for this experimentally simpler configuration are similar to the ones in {\bf (a)}.}
\end{figure}

 In Fig.~\ref{fig:6 ion chain}{\bf (a)}, we show the optimal fidelity for the dissipative generation of the ideal
ground state at different fillings~\eqref{gs_cooling}, by integrating numerically the dipolar and inhomogeneous
trapped-ion DSBC model~\eqref{dsbc_ion} as a function of the
ratio between heating and cooling rate $\zeta$. We study a chain of $N+2=6$ ions, in which $N=4$ have a hyperfine
structure and play the role of the spins in the XY chain, while the two peripheral
ions are assumed to be used for sympathetic cooling (this choice of auxiliary cooling ions at the ends of
the chain makes the interparticle spacing slightly more homogeneous for the ``system'' ions). As can be seen in the
figure, when $\zeta\to0$,
the fidelities with the target ground state approach 100$\%$, which allows us to conclude that the differences due
to the inhomogeneity of the chain and the dipolar range of interactions with respect to the ideal homogeneous and
nearest-neighbor DSBC~\eqref{dsbc} do not compromise the fidelities severely. 

 The situation is different as finite heating rates $\zeta>0$ are considered. In Fig.~\ref{fig:6 ion chain}{\bf (a)}, we
observe that the  fidelity with the target ground state decays essentially linear with $\zeta$. In the worst case
considered, where $\zeta=0.2$ leads to a mean phonon number of $\bar{n}_{1z}\simeq0.25$, the fidelities obtained are
reduced
to roughly $\mathcal{F}_{t_{\rm f}}\approx 0.8$. We note that resolved
sideband cooling of single vibrational modes has been experimentally achieved reaching $\bar{n}_{1z}$ below 0.1 for
single trapped ions~\cite{leibfried_review}. Sympathetic cooling of crystals up to $N=4$ ions has also been
achieved~\cite{sympathetic_cooling}, reaching values as low as $\bar{n}_{1z}\approx0.06$ with
pulsed techniques. We note
that our scheme does not demand such an accurate ground-state cooling (see Fig.~\ref{fig:6 ion 
chain}{\bf (a)}), and we expect that the required mean phonon numbers $\bar{n}_{1z}\approx0.1$ can also be
achieved in a continuous cooling scheme.

Another source of experimental imperfections is given by slow drifts of the trap frequencies which will modify
the detuning $\delta_{1z}$ in different experimental runs, such that the conditions~\eqref{conditions_down} will not be
perfectly fulfilled. Nonetheless, in Fig.~\ref{fig:fidelity_N3} we have shown that the dissipative character of the
protocol endows it with a natural robustness with respect to non-optimal choices of the parameters. Therefore, one can
expect large fidelities even in the presence of slow drifts of the trap frequency within the kHz-range. In any case, to
alleviate imperfections caused by both magnetic-field dephasing and trap frequency drifts, we have considered a faster
protocol where the required time is $\tilde{t}_{\rm f}\sim 10^2/J\approx$1-10$\,$ms. In Fig.~\ref{fig:6 ion chain}{\bf
(b)}, we show that the achieved fidelities in this case are still above 0.8 (for heating/cooling ratios below 0.2).
Moreover, in this figure we have also
considered substituting the standing-wave force in Eq.~\eqref{spin_boson_ion} by a traveling-wave configuration, which
is experimentally less demanding. 

The need of the $\tau^z$ state-dependent
force~\eqref{z_force} forbids the use of magnetic-field insensitive states (i.e. clock states). Therefore, the hyperfine
spins will be subject to fluctuations induced by non-shielded external magnetic fields, typically leading to
dephasing in a timescale of $T_2\approx$1-10$\,$ms. These timescales are comparable to the protocol times $t_{\rm
f}\sim 10^2/J\approx$1-10$\,$ms, where we have taken $J/2\pi\sim$1-10$\,$kHz. Nevertheless,  these magnetic-field
fluctuations act globally for standard radio-frequency traps $H_{\rm n}=\half\Delta\omega_0(t)\sum_i\tau_i^z$, where
$\Delta\omega_0(t)$ is a fluctuation of the resonance frequency due to the Zeeman shift (note that this might not be the
case for micro-fabricated surface traps, where fluctuating magnetic-field gradients can also arise). Since the dynamics
of the DSBC~\eqref{dsbc_ion} conserves the number of spin excitations, this 
magnetic-field noise only introduces a global fluctuating phase, and thus does not decohere the state of the system. 
Therefore, as far as the induced spin dynamics occurs within any subspace with a conserved number of excitations, the
dissipative protocol is robust to global magnetic-field noise. 

\begin{figure}
\centering
 \includegraphics[width=0.9\columnwidth]{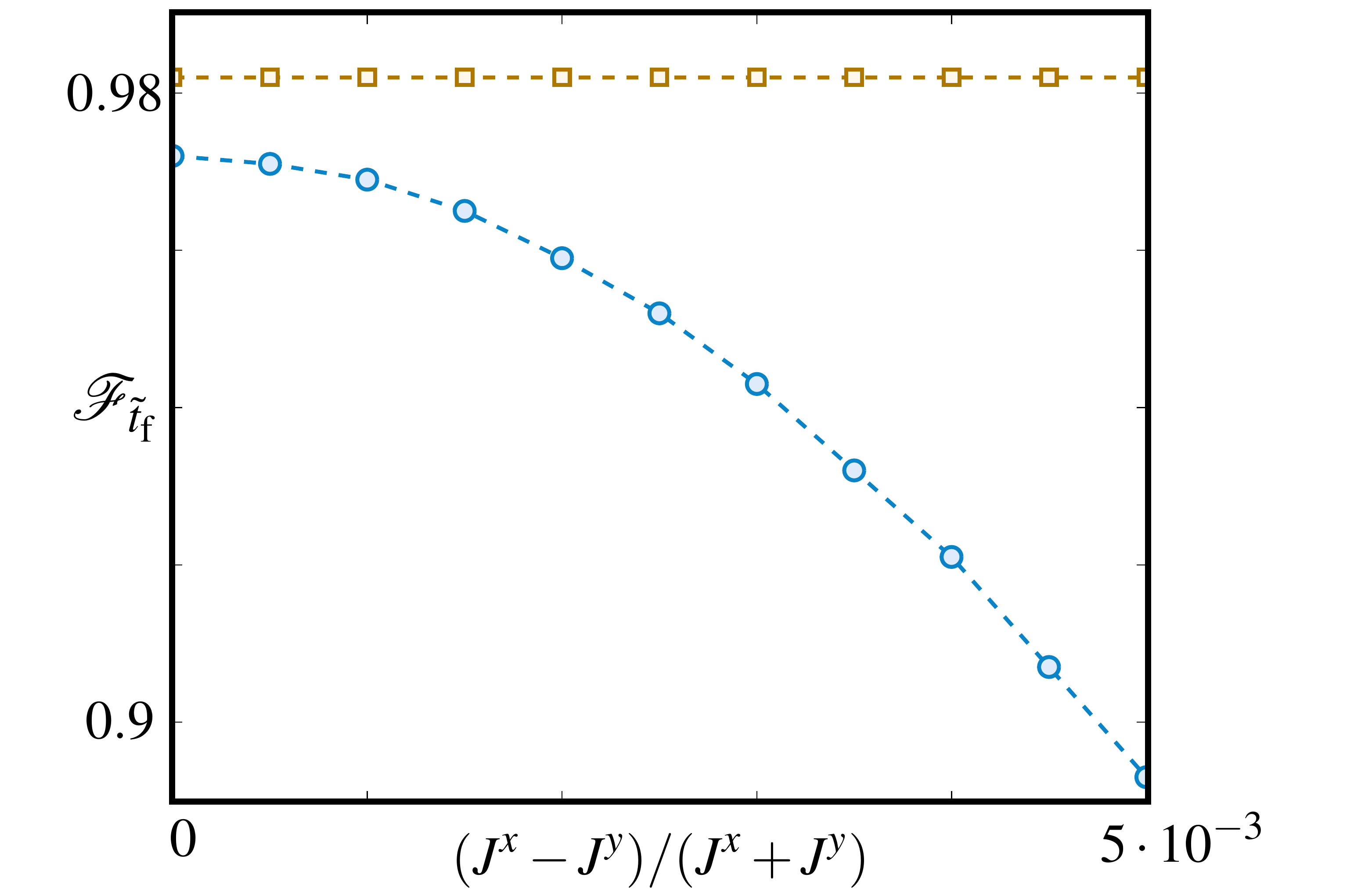}
\caption{    
\label{fig:7 ion chain} {\bf Anisotropy in the trapped-ion dissipative protocol:} {\bf (a)} Fidelity with the desired
target state  for a
fixed
time $t_{\rm f}=10^2/J$ and as a function of the anisotropy ratio $(J^x-J^y)/(J^x+J^y)$  for a chain of six ions
in which the central four ions represent sites of the slightly-anisotropic XY spin chain. The blue dots correspond to
the case of
$n_{\rm s}=\{1,3\}$ initial spin excitations, and the orange squares to the case of $n_{\rm s}=2$ excitations. The
oscillator was
truncated to five levels.}
\end{figure}

It is, however, important to stabilize the intensity of the
state-dependent forces to the sweet spot where $J_{ij}^x=J_{ij}^y$, and make sure that
$|\mathcal{F}^{\alpha}_{ij}|\ll\delta_{n\alpha}$ to neglect any term that does not conserve the number of excitations.
The
effect of off-resonant terms in the scheme leading to the XY interactions, that could give rise to non-conservation of
the number of spin
excitations, should be much smaller than in stroboscopic proposals \cite{digital_dissipation_ions} since our protocol
does not require fast gates \cite{Kirchmair_NJP_2009}. In Fig.~\ref{fig:7 ion chain}, we show the achieved fidelities
for anisotropies in
$J_{ij}^x/J_{ij}^y\neq 1$. The results have been obtained numerically for the same regime as in Fig.~\ref{fig:6 ion
chain}{\bf (b)}, and indicate that if $J_{ij}^x,J_{ij}^y$ differ by 1 part in $10^3$, the fidelities are not severely
reduced. The results in Fig.~\ref{fig:7 ion chain} also show a marked difference in the robustness depending on the
number of spin excitations in the initial state. This behaviour, at first surprising, can be understood by looking at
the small
undesired Hamiltonian terms responsible for the anisotropy in an interaction picture with respect to the ideal isotropic
Hamiltonian. The anisotropy term creates or destroys fermionic quasiparticles in pairs, and rotates with a frequency
that is the sum of the energies of the two quasiparticles. For the case with only one spin up, it is possible to create
from the ideal target state pairs of fermions with energy adding up to zero, so that these error terms do not rotate. In
the case with two spin excitations, on the contrary, the lower half of the fermionic spectrum is filled, so that any
Hamiltonian term creating or destroying a pair of fermions from the ideal target state oscillates in time, and therefore
its effect is strongly suppressed.

\bigskip

\section{Conclusions and Outlook}
\label{conclusions}

We have presented a method to dissipatively generate multipartite-entangled states corresponding to the ground
states of small spin chains with isotropic XY Hamiltonian in a transverse field. The protocol for dissipative state
preparation is interesting from a fundamental point of view, as it illustrates
how local and even purely Markovian noise can assist entanglement production. Indeed, the jump operators required are
sums of
terms acting on individual sites, in contrast with the few-body (quasi-local) nature of the operators in
\cite{dissipative_pure_state_engineering, driven_dissipative_condensate}.
When the noise deviates from exact Markovianity, the required time for achieving the steady state is reduced, which
illustrates the value of non-Markovian effects for practical purposes.
We would like to emphasize that the general
idea underlying the procedure presented in this work is not specific to the model considered, and might be generalized
to other
spin Hamiltonians. As an example, we introduced a variation that can be used to generate states locally equivalent to
W-states. It would also be interesting to modify our dissipative protocol to prepare the ground state of a gapped spin
model. For longer ion chains, the combination of more state-dependent forces~\eqref{spin_boson}
with different wavevectors could improve the scalability of the protocol. 

We have also explained in detail how to implement this method in small chains of trapped ions. In
our proposal, two internal levels of the ions embody the spin system and a collective motional mode represents the
damped oscillator. The preparation procedure
requires the implementation of spin-spin interactions using the so-called M{\o}lmer-S{\o}rensen scheme, the action of a state-dependent
force, and sympathetic cooling, all ingredients within the capabilities of present ion-trap technology. 
Numerical simulations including different sources of errors indicate that the protocol can produce the target
states with fidelities comparable to the more standard coherent protocols~\cite{entangled_states_ions}. The method presented
is indeed robust against a number of experimental imperfections, and the
implementation is simplified since only global addressing of the ions is required. 

Our results, however, may find a realization in different experimental setups. For instance, the application of
our ideas to arrays of superconducting qubits in stripline resonators seems feasible. Morever, since this model can
also be understood as a
quadratic fermonic model with an additional chemical potential fixing the number of particles, the results may also be interesting for fermionic atoms in optical lattices.

The realization of this kind of dissipative system paves the way for the implementation of more complex scenarios
to study the interplay of coherent and incoherent dynamics giving rise to noise-induced criticality
\cite{noise_crit}. Indeed, recent work in the area of non-equilibrium quantum phase
transitions \cite{non-eq_qpt} has shown a number of fascinating results that could be demonstrated in systems of trapped
ions using tools similar to the ones in our protocol. 

{\it Acknowledgements.--}  We thank A. Rivas for fruitful comments on the manuscript. This work was supported by  PICC
and by the Alexander von Humboldt Foundation. A.B. thanks FIS2009-10061, and QUITEMAD.


\end{document}